\documentclass[twocolumn]{aastex631}

\usepackage{url}

\usepackage{amsmath}
\usepackage{float}
\usepackage{multirow}
\usepackage{soul}

\usepackage{fontawesome}   
\usepackage{hyperref}      
\usepackage{xcolor}

\definecolor{linkcolor}{rgb}{0.1216,0.4667,0.7059}
\newcommand{\codelink}{%
  \href{https://github.com/wangxianyu7/Data_and_code/blob/main/UnifiedKraftBreak/Find_Kraft_Break.ipynb}%
       {{\color{linkcolor}\faFileCodeO}}\,\,}
\newcommand{\oscaption}[1]{\caption{#1 \codelink}}

\newcommand{\feh}{\ensuremath{\left[{\rm Fe}/{\rm H}\right]}}

\newcommand{\teff}{\ensuremath{T_{\rm eff}}}

\newcommand{\logg}{\ensuremath{\log g_\star}}

\newcommand{\msun}{\ensuremath{\,M_\Sun}}

\newcommand{\mstar}{\ensuremath{M_{\star}}}

\newcommand{\ar}{\ensuremath{a/R_\star}}
\newcommand{\vsini}{\ensuremath{v\sin{i_\star}}}

\newcommand{\degrees}{\ensuremath{^{\circ}}}
\renewcommand{\bv}{Brunt-V\"ais\"al\"a}

\newcommand{\singlelambdacut}{{6447^{+85}_{-119}}}
\newcommand{\singlelambdacutvalue}{{6447}}
\newcommand{\binarylambdacut}{{6105^{+123}_{-133}}}
\newcommand{\binarylambdacutvalue}{{6105}}
\newcommand{\singlevsinicut}{{6510^{+97}_{-127}}}

    {\clearpage\pagebreak[4]\global\pdfpageattr\expandafter{\the\pdfpageattr/Rotate 90}}%
    {\clearpage\pagebreak[4]\global\pdfpageattr\expandafter{\the\pdfpageattr/Rotate 0}}%

\usepackage{CJKutf8}

\begin{document}

\title{Unified Kraft Break at $\sim\,$6500\,K: \\
A Newly Identified Single-Star Obliquity Transition Matches the Classical Rotation Break}

 \author[0000-0002-0376-6365]{Xian-Yu Wang {\begin{CJK}{UTF8}{gbsn}(汪宪钰)\end{CJK}}}
 \altaffiliation{Sullivan Prize Postdoctoral Fellow}
 \affiliation{Department of Astronomy, Indiana University, 727 East 3rd Street, Bloomington, IN 47405-7105, USA}

 \author[0000-0002-7846-6981]{Songhu Wang {\begin{CJK}{UTF8}{gbsn}(王松虎)\end{CJK}}}
 \affiliation{Department of Astronomy, Indiana University, 727 East 3rd Street, Bloomington, IN 47405-7105, USA}

 \author[0000-0001-7664-648X]{J. M. Joel Ong {\begin{CJK}{UTF8}{gbsn}(王加冕)\end{CJK}}}
 \affiliation{Sydney Institute for Astronomy (SIfA), School of Physics, University of Sydney, NSW 2006, Australia}
 \affiliation{Institute for Astronomy, University of Hawai`i, 2680 Woodlawn Drive, Honolulu, HI 96822, USA}

 \correspondingauthor{Songhu Wang}
 \email{sw121@iu.edu}
\begin{abstract}

The {stellar obliquity} transition, defined by a $T_{\rm eff}$ cut separating aligned from misaligned hot Jupiter systems, has long been assumed to coincide with the rotational Kraft break. Yet the commonly quoted obliquity transition (6100 or 6250~K) sits a few hundred kelvin cooler than the rotational break ($\sim$\,6500~K), posing a fundamental inconsistency. We show this offset arises primarily from \textit{binaries/multiple-star systems}, which drive the cooler {stellar obliquity} transition ($\binarylambdacut$~K), although the underlying cause remains ambiguous. After removing binaries and higher-order multiples, the single-star {stellar obliquity} transition shifts upward to $\singlelambdacut$~K, in excellent agreement with the single-star rotation break ($\singlevsinicut$~K). This revision has two immediate consequences for understanding the origin and evolution of spin-orbit misalignment. First, the upward shift reclassifies some hosts previously labeled `hot' into the cooler regime; consequently, there are \emph{very} few RM measurements of non-hot-Jupiter planets around genuinely hot stars
 ($\teff\gtrsim6500\,\mathrm{K}$), and {previously} reported alignment trends for {these classes of systems (e.g., warm Jupiters and compact multi-planet systems)} lose the power to discriminate the central question: are large misalignments unique to hot-Jupiter-like planets that can be delivered by high-$e$ migration, or are hot stars intrinsically more misaligned across architectures? Second, a single-star {stellar obliquity} transition near $6500\,\mathrm{K}$, coincident with the rotational break, favors tidal dissipation in outer convective envelopes; as these envelopes thin with increasing $\teff$, inertial-wave damping and magnetic braking weaken in tandem.
\end{abstract}


\section{Introduction} \label{sec:intro}

The mass of a star largely determines how it \emph{generates} energy, and how this energy is \emph{transported}. Relatively low core temperatures in low-mass cool dwarfs ($0.5 - 1.3\,\msun$) favor hydrogen burning through the proton-proton chain, with energy carried outward through a deep convective envelope. In hotter, more massive dwarfs, the CNO cycle dominates hydrogen burning in a convective core, while the overlying envelope is predominantly radiative. These structural differences have been supported by multiple observational tracers. For example, cool dwarfs show envelope-driven mixing that depletes surface lithium \citep{Boesgaard1986, Hobbs1986, Deliyannis1997, Sestito2005}. Where asteroseismology is observationally available, this convection stochastically excites solar-like oscillations \citep{Kjeldsen1995,Chaplin2013,Silva2013}. Hotter F stars, with shallower convective envelopes, exhibit far shorter convective turnover and damping timescales in their pulsations. Stars which are hotter still possess negligible convective envelopes, and display only classical pulsations (e.g. $\delta$~Scuti/roAp) driven by the $\kappa$ mechanism \citep{Cox1980}. 

A much more easily accessible observational manifestation of this structural shift is the sharp drop in rotation rates at mid-F spectral types: the well-known \emph{Kraft break}\footnote{Explicit evidence for a sharp drop in rotation near F5 long predates Kraft’s work (e.g., \citealt{Struve1931,Westgate1934}).} \citep{kraft1967break}. A sufficiently thick convective envelope in cool stars sustains a global magnetic dynamo, whose efficiency is described by the dimensionless Rossby number, operationally defined as the ratio between the rotational period and the convective turnover timescale \citep{noyes_rotation_1984,brun_magnetism_2017}. Low Rossby numbers in cool stars are generally associated with higher magnetic activity, manifesting as enhanced spots, frequent flares, and strong chromospheric/coronal emission \citep[e.g.][]{saar_time_1999,pizzolato_stellar_2003,mamajek_improved_2008,reiners_evidence_2009,vidotto_stellar_2014,stelzer_path_2016,newton_halpha_2017, Feinstein2020, see_further_2023,mathur_magnetic_2025}. Magnetized stellar winds extract angular momentum and {slow the rotation of stars as they age} \citep[e.g.][]{weber_AM_1967,skumanich_timescales_1972,kawaler_rotational_1989,barnes_gyrochronology_2003,vansaders_fast_2013}. By contrast, with thinner convective envelopes, magnetic braking is much less efficient in hot stars, which remain rapidly rotating on the main sequence.

{Independently, exoplanet observations have revealed a transition in the stellar spin-orbit angle, the angle between the stellar rotation axis and the planetary orbital axis (hereafter obliquity), among host stars:} hot Jupiters orbiting hotter stars more likely exhibit large misalignments, whereas those around cooler stars are usually well aligned \citep[e.g.,][]{Schlaufman2010, Winn2010, Albrecht2012, Albrecht2021, Knudstrup2024}. This is often attributed to efficient tidal obliquity damping \citep{Albrecht2012}, via dissipation of tidally excited inertial waves within convective envelopes \citep{OgilvieANDLin2007, Lai2012, Li2016, saunders_efficient_2024}, and, where relevant, by resonances between the dynamical tides and stellar oscillations \citep{zanazzi2024damping,Zanazzi2025}, whereas such damping is inefficient in hot, radiative-envelope stars.

{Because both phenomena reflect a structural break separating cool and hot stars, it has long been assumed that the \emph{rotation} and \emph{obliquity} transitions coincide. Whether they do has remained unclear. Historically, both the rotation \citep[e.g.,][]{kraft1967break,noyes_rotation_1984,vansaders_fast_2013,Avallone2022} and obliquity \citep[e.g.,][]{Winn2010,Albrecht2012, WinnFabrycky2015, Triaud2018, Albrecht2022, Knudstrup2024} literature have been content to adopt notional thresholds, treating their locations as well established, despite rarely determining them through formal statistical fitting and, in many cases, without identifying the boundary explicitly in terms of stellar effective temperature. As a result, the quoted transition locations vary across studies, leading to somewhat disparate breaks: a mid-F (i.e. $M_\star=1.2$-$1.3\,\msun$) divider, or the equivalent in color or temperature (such as $B\!-\!V=0.3$, $G_{\rm BP}\!-\!G_{\rm RP}=0.55$-0.60, or $\teff \sim 6500\ \mathrm{K}$: \citealt{Iben1967Comparison_of_Theory_with_Observation,donascinmento_lithium_2000,vansaders_forward_2019}) for rotation, and $\teff=6000$-$6300$~K (often 6100 or 6250~K, {as reviewed by \citealt{Albrecht2022}}) for obliquity, without a formal turning-point inference or quoted uncertainties.}

Recent work has sharpened one side of this picture. Using a carefully curated sample of nearby single F-dwarfs, \citet{Beyer2024} found a sharp rotation transition centered at $\teff=6550$~K (width of 200~K), arguing that the commonly cited $6100$~K obliquity transition is $450$~K cooler than this value. Although the obliquity sample (dominated by hot-Jupiter hosts) and the rotation sample (composed of F stars within 33~pc) do differ in \feh\ (medians $0.04^{+0.20}_{-0.16}$ vs.\ $-0.15^{+0.19}_{-0.18}$ ) --- which can affect the Kraft-break location by altering envelope opacity (and thus convective-zone depth and magnetic-braking efficiency) --- a $\sim\!0.2$\,dex metalicity offset is expected to shift the break only by less than $100$~K \citep{Spalding2022}: far smaller than the observed $450$~K gap. 

We notice that the rotation result pertains to a rigorously vetted sample of \emph{single} stars. The often-cited obliquity divider near $6100$~K, however, is typically drawn from literature samples that include binaries and multiple-star systems. As a result, a direct apples-to-apples comparison between the rotation and obliquity breaks in single-star systems does not yet exist.

In this work, we supply one: we present a consistent statistical determination of the transition points in both stellar $\vsini$ (rotation) and $\lambda$ (obliquity), enabling a rigorous comparison of like with like. We apply a uniform “turning-point’’ fitting methodology to an updated, single-star rotation sample based on \citet{Beyer2024}, and confirm a sharp rotation transition at $\teff=\singlevsinicut$~K. Applying the same procedure to a hot-Jupiter Rossiter-McLaughlin (RM, \citealt{Rossiter1924,McLaughlin1924}) sample, we recover an obliquity transition consistent with commonly quoted values, $\teff=\binarylambdacut $~K. Thus, our baseline analysis reproduces a similar but slightly smaller $405$~K  offset to that noted by \citet{Beyer2024}. Crucially, when we remove binaries from the obliquity sample so that both probes refer to \emph{single} stars, the inferred obliquity break shifts upward to $\teff=\singlelambdacut$~K, which is statistically indistinguishable from the rotation break ($\singlevsinicut$~K), differing by only $0.41\,\sigma$. That is to say, 
a \emph{single} mid-F boundary at $\sim 6500$~K consistently describes both the rotation (Kraft) break and the spin-orbit alignment transition, once binarity is controlled for.

\section{Sample construction}\label{sec:sampleconstruction}

\subsection{Stellar Rotation, \vsini}\label{sec:sampleconstruction:vsini}

\begin{figure}
    \centering
    \includegraphics[width=1\linewidth]{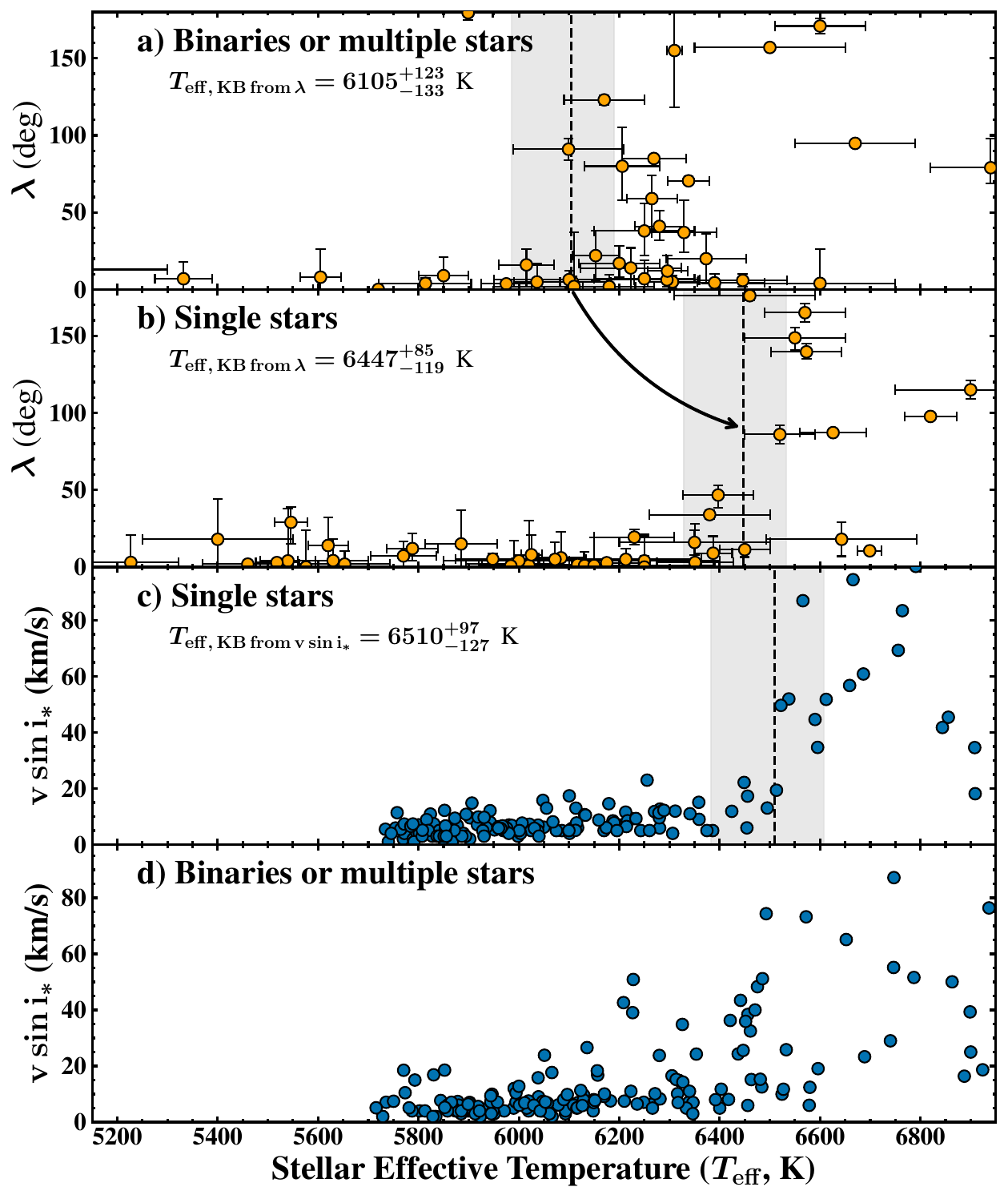}
    \oscaption{The $\lambda$ and \vsini\, distributions, along with derived \teff\, boundaries. \textit{Panel a:} $\lambda$ distribution for hot-Jupiter systems with {confirmed or candidate stellar companions}. The \teff\, boundary of $\binarylambdacut$~K is marked with a black dashed line, with shaded 1 $\sigma$ interval. \textit{Panel b:} Same as the top panel, but for single-star hot-Jupiter systems with \teff\, boundary = $\singlelambdacut$ K. \textit{Panel c:} \vsini\, distribution for single stars, compiled following the {updated} prescription described in \citet{Beyer2024}. The resulting \teff\, boundary is $\singlevsinicut$~K, consistent with that derived from the single-star $\lambda$ distribution. \textit{Panel d:} same as panel c, but for binary and multiple-star systems.}
    \label{fig:kb}
\end{figure}

To statistically identify the turning point of the Kraft rotation break, we assembled a sample of 405 systems with \teff\ and $\vsini$ measurements (panels c and d of \autoref{fig:kb}), following the general selection and vetting criteria described in Section 2 of \citet{Beyer2024}, but with minor modifications aimed at expanding the sample and minimizing the influence of binary/multiple-star systems. 

We compiled all bright F-type stars within 33.33 pc of the Sun that have reliable parallax measurements, using the \texttt{ADQL} query described in Section 2 of \citet{Beyer2024}. The primary source of \teff\, is \texttt{teff\_gspphot}; if unavailable, we adopt \texttt{teff\_gsspec} or \texttt{teff\_esphs}\footnote{\url{https://gaia.aip.de/metadata/gaiadr3/astrophysical_parameters/}}. If none of these are available, \teff\, values from the Geneva–Copenhagen Survey \citep{Nordstrom2004} and the stellar property catalog of rapidly rotating Sun-like stars by \citet{Schroder2009} were supplemented. {As for} \vsini\, the Gaia DR3 line-broadening measurements (\texttt{vbroad}) were adopted as a proxy \citep{Fremat2023}. When \texttt{vbroad} was unavailable, we instead used \vsini\ values from \citet{Nordstrom2004} and \citet{Schroder2009}. Following \citet{Beyer2024}, we removed young stars (with ages less than 200 Myr) and evolved stars (with $\logg < 3.75$), as these may exhibit anomalous rotation rates. We identified systems with stellar companions via multiple steps. Firstly, binaries and binary candidates were identified by cross-matching our targets with the Geneva–Copenhagen Survey \citep{Nordstrom2004}, where entries marked with an asterisk in the \texttt{$f_d$} (confirmed or suspected binaries) or \texttt{$f_b$} (spectroscopic binaries) flags were considered binary or potential binary systems. Next, we identified stellar companions following the methods described in \citet{Badry2021}, but without the $G < 18$ limitation in order to include faint stellar companions. Finally, stars with Renormalized Unit Weight Error (RUWE, \citealt{Lindegren2018,Lindegren2021}) $>$ 1.4 were classified as potential binary or multiple-star systems. The final sample comprises 405 stars with \teff\ and \vsini\ measurements, including 182 mature, single main-sequence stars, 206 binary or multiple-star systems (see panels c and d of \autoref{fig:kb}), and 17 young or evolved stars, which are excluded from our analysis.

\vspace{1cm}
\subsection{Stellar Obliquity, $\lambda$}\label{sec:sampleconstruction:stellarobliquity}

To determine the boundary of the obliquity transition, we focus on systems that would be affected by tidal realignment, if operative. Concretely, we select hot Jupiters with mass ratios
$3\times10^{-4} \leq m_{\rm p}/M_\star \leq 2\times10^{-3}$ \citep{Rusznak2025} on close-in orbits with $a/R_\star < 10$, and with measured stellar obliquities. Stellar masses, planetary masses, and $a/R_\star$ are taken from the Planetary Systems Composite Parameters table (\texttt{PSCompPars}\footnote{\url{https://exoplanetarchive.ipac.caltech.edu/cgi-bin/TblView/nph-tblView?app=ExoTbls&config=PSCompPars}}, \citealt{10.26133/NEA2}) at NASA Exoplanet Archive \citep{Christiansen2025}; when a value is missing there, we adopt the entry from the Encyclopaedia of exoplanetary systems\footnote{\url{https://exoplanet.eu/catalog/}}  as a fallback. 

Among stellar-obliquity measurement techniques, the Rossiter-McLaughlin method is the most widely used and provides the majority of $\lambda$ measurements \citep{WinnFabrycky2015, Triaud2018, Albrecht2022}. Because different techniques introduce different selection effects (e.g., spot-crossing is more sensitive to aligned systems, while gravity-darkening is usually significant only for highly misaligned systems, \citealt{Siegel2023,Albrecht2022}), we retain only RM-based measurements, giving priority in the following order: classical RM, Doppler shadow \citep{Albrecht2007, Collier2010, Zhou2016, Johnson2017}, reloaded RM \citep{Cegla2016ReloadedRM}, and RM Revolution \citep{Bourrier2021RMrevolutions}. For all retained systems, we adopt the sky-projected obliquity $\lambda$ and \teff\, from \texttt{TEPCat}\footnote{\url{https://www.astro.keele.ac.uk/jkt/tepcat/obliquity.html}, \\accessed on 2025~October~04. } \citep{Southworth2011}.

\begin{figure}
    \centering
    \includegraphics[width=1\linewidth]{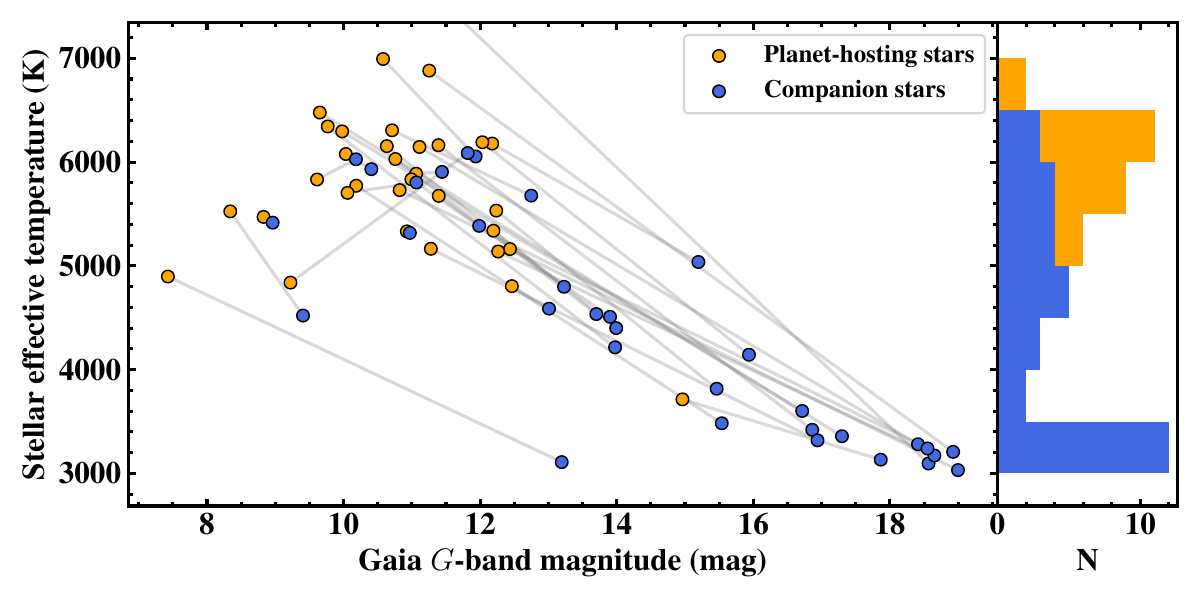}
    \caption{Stellar effective temperature (\teff) and Gaia DR3 $G$ magnitude for the stellar obliquity sample with stellar companions. Planet-host stars and their corresponding stellar companions are shown as orange and blue dots, respectively. The gray lines connect each host-companion pair. The \teff\ distributions of the two populations are shown in the right panel. {Interestingly, in our stellar-obliquity sample (consisting of hot Jupiters), the planet host is \emph{always} the more massive component of a binary.}}
    \label{fig:teffgmag}
\end{figure}

{
Following \citet{Albrecht2022}, we exclude low-precision measurements with $\sigma_\lambda > 50$\degrees\, and controversial cases flagged in the literature: CoRoT-1 (\citealt{Bouchy2009CoRoT1} reported an aligned configuration, while \citealt{Pont2010} found a significant misalignment of $\lambda = 77 \pm 11$\degrees), CoRoT-19 \citep[][the RM effect was detected at only a 2.3$\sigma$ confidence level]{Guenther2012}, HATS-14 \citep[][the $\lambda$ constraint is sensitive to the \vsini\, prior and RM offset modeling due to the lack of post-egress data]{Zhou2015}, HAT-P-27 \citep[][large $\lambda$ uncertainty: $\lambda = 24.2^{+76.0}_{-44.5}$\degrees]{Brown2012}, HD 3167 (a large mutual inclination of $\sim 100$\degrees\, was inferred from literature $\lambda$ measurements, \citealt{{Dalal2019}} and \citealt{Bourrier2021RMrevolutions}, although \citealt{Teng2025} showed that such a configuration is unlikely to be maintained), WASP-1 \citep[][a weak signal of a prograde orbit was detected with $\sim2\sigma$ confidence]{Simpson2011a,Albrecht2011}, WASP-2 \citep[][no RM signal was detected]{Triaud2010,Albrecht2011}, WASP-23 \citep[][the $\lambda$ constraint is sensitive to the $v\sin i$ prior due to the low transit impact parameter]{Triaud2011}, WASP-49 \citep[][large $\lambda$ uncertainty: $\lambda = 54^{+79}_{-58}$\degrees]{Wyttenbach2017}, WASP-134 \citep[][the $\lambda$ constraint is sensitive to assumptions on priors and RM offset due to the lack of post-egress data]{Anderson2018a}, and HAT-P-17 (an aligned configuration of $\lambda = 19^{+14}_{-16}$\degrees was initially reported by \citealt{Fulton2013}, but a misalignment of $\lambda = -27.5 \pm 6.7$\degrees was later found by \citealt{Mancini2022})
}

Stellar multiplicity has long been theorized as a potential driver of spin-orbit misalignments \citep{Holman1997, Wu2003, Fabrycky2007, Naoz2012, Batygin2012}. Recent obliquity studies confirm that warm Jupiters tend to be aligned in single-star systems \citep{Rice2022WJs_Aligned, WangX2024,Espinoza2023} yet are frequently (and sometimes famously, e.g., HD~80606; \citealt{Winn2009}; TIC~241249530; \citealt{Gupta2024}) misaligned in binaries; massive planets show a similar single-versus-binary contrast \citep{Rusznak2025}. Motivated by this, we identify and flag binaries or higher-order multiples in our {obliquity} sample using \citet{Schwarz2016} and \citet{Fontanive2021}. Similar to \autoref{sec:sampleconstruction:vsini}, we examined the binarity of each system following the method of \citet{Badry2021}, but extended it to include faint nearby stars ($G > 18$). We also classified systems with RUWE $>$ 1.4 as binary candidates. Moreover, we only include systems with stellar ages $>$ 200 Myr and $\logg$ $>$ 3.75 to exclude young and evolved stars, which is consistent with the construction of the \vsini\, sample.

After excluding poor or controversial obliquity measurements, as well as young and evolved stars, the hot-Jupiter obliquity samples used for our turning-point analysis consist of 91 systems in total, of which 50 are single-star systems (See panels a and b of \autoref{fig:kb}).

\begin{figure}
    \centering
    \includegraphics[width=1\linewidth]{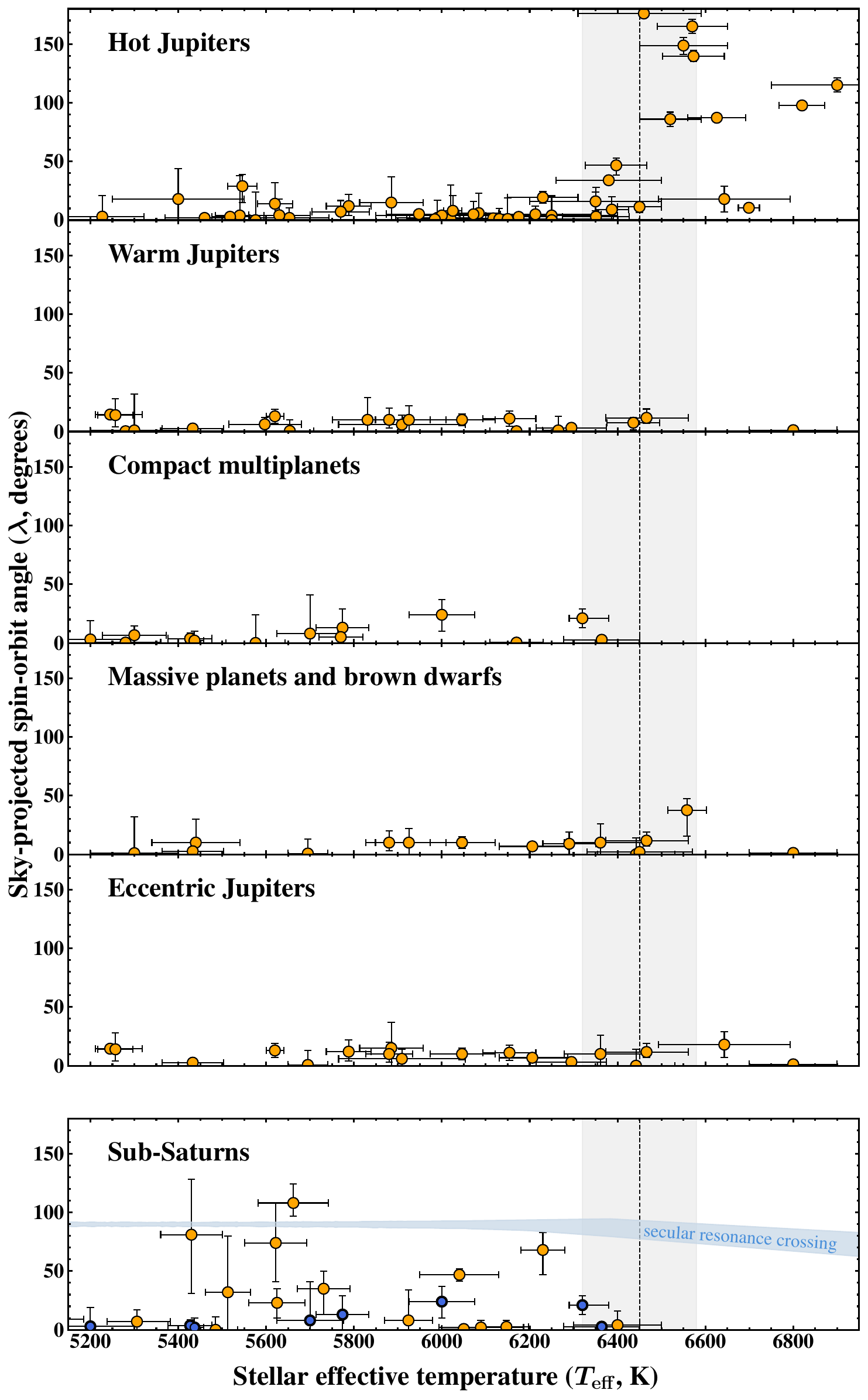}
    \caption{{The $\lambda$ distributions for hot-Jupiter ($3\times10^{-4} \leq m_{\rm p}/M_\star \leq 2\times10^{-3}$, \ar$<$10), sub-Saturn ($3\times10^{-5} \leq m_{\rm p}/M_\star \leq 3\times10^{-3}$), warm-Jupiter ($3\times10^{-4} \leq m_{\rm p}/M_\star \leq 2\times10^{-3}$, \ar$>$11), compact multiplanet (period ratio of adjacent planets, $P_{i+1}/P_{i}<6$), massive-planet systems ($2\times10^{-3} \leq m_{\rm p}/M_\star$), and eccentric Jupiters ($3\times10^{-4} \leq m_{\rm p}/M_\star \leq 2\times10^{-3}$, $e>0.1$),} along with the Kraft break and its associated uncertainty derived in this work. In terms of \teff\, coverage, only the hot-Jupiter systems have sufficient measurements ($>10$) in the hot-star regime. Sub-Saturns with nearby companions are marked in blue. The predicted relation between \teff\, and $\lambda$ from secular resonance crossing \citep{Petrovich2020,Dugan2025} is shown as the light-blue shaded region.}
    \label{fig:trends}
\end{figure}

\vspace{1cm}
\section{\teff\, Boundary Determination}\label{sec:stat}

To identify the \teff\, boundary that best separates cool (low \vsini, aligned) and hot (high \vsini, misaligned) systems, we evaluated candidate cuts at the midpoints between adjacent unique \teff\, values in each statistical test.

For each candidate cut, we required at least five systems in both the cool and hot populations to ensure adequate sample sizes. We then compared the $\lambda$ or \vsini\, distributions of the two populations using the two-sample Kolmogorov-Smirnov (KS; \citealt{Hodges1958TheSP}) test as implemented in \texttt{scipy} \citep{virtanen2020scipy}. We quantified the statistical significance as $-\log_{10}(p)$, where $p$ is the KS test $p$-value, since the logarithmic scale facilitates comparison across small $p$-values. We defined the optimal boundary as the \teff\, cut that maximized this quantity.

To estimate the uncertainty in the optimal boundary, we performed a bootstrap analysis. We generated $100{,}000$ bootstrap realizations by resampling the targets with replacement. In each realization, we perturbed every star’s $\teff$ by drawing from a Gaussian distribution centered on its reported value with standard deviation equal to its measurement uncertainty (i.e., $\sigma_{\teff}$). Note that a conservative 2.4\% floor uncertainty in \teff\ was adopted \citep{Tayar2022}. For each bootstrap sample, we repeated the boundary search over the same $\teff$ grid and identified the turning point that maximized the two-sample KS statistic. The resulting distribution of $100{,}000$ turning points was then used to derive the cut and the corresponding lower and upper uncertainties.

We applied this procedure to the stellar-obliquity distribution of hot-Jupiter systems whose hosts had stellar companions (\autoref{fig:kb}, panel a) and obtained a boundary at $\binarylambdacut$\,K\footnote{We exclude KELT-23 A, a strongly misaligned outlier, and the four $T_{\rm eff}\geq6600$\,K hosts, which lie far from the boundary.}, consistent with the commonly quoted $6100-6250$\,K range, confirming that the previously reported obliquity transition \teff\ is driven by hot-Jupiter systems in binaries. For hot Jupiters around single stars (\autoref{fig:kb}, panel b), the turning point of their obliquity distribution is $\singlelambdacut$\,K, in excellent agreement with the \vsini\ break for single stars that we measure at $\singlevsinicut$\,K (\autoref{fig:kb}, panel c).

\section{Why the Break Looks Cooler in Binary Samples}
\label{sec:binaries}

Our statistical analysis shows that, when we restrict the sample to \textit{single stars}, the Kraft rotation break and the obliquity transition coincide near $\teff \sim 6500\,\mathrm{K}$ (see \autoref{fig:kb}, panels b and c). In contrast, for \textit{binaries and higher-order multiples}, the misaligned fraction rises sharply above $\teff \sim 6100\,\mathrm{K}$, matching the often-quoted threshold that our results suggest was influenced by the inclusion of multiple systems (see \autoref{fig:kb}, panel a). Interestingly, in the multiple-star sample, the rotation break also trends toward $6100\,\mathrm{K}$ rather than $6500\,\mathrm{K}$ (see \autoref{fig:kb}, panel d), although there is no single preferred rotational boundary in the binary sample: $-\log_{10}(p)$ remains near its maximum throughout $6100–6500\,\mathrm{K}$. Taken together, these patterns indicate that stellar companions either affect spin-orbit misalignment and alter rotational evolution in a coupled fashion, or bias the inferred $\teff$.

As noted in \autoref{sec:sampleconstruction:stellarobliquity}, both theoretical models and observational evidence demonstrate that stellar companions \emph{can} efficiently tilt planetary systems. {Assuming a representative protoplanetary disk with a disk mass of $m_{\rm disk}=10^{-2}\mstar$, an inner edge at $a_{\rm in}=0.05\,\mathrm{AU}$ and an outer edge at $a_{\rm out}=50\,\mathrm{AU}$ following a surface density profile $\Sigma \propto r^{-1}$, together with an eccentricity of $e' = 0.5$ and a mutual inclination of $i' = 45\degrees$ between the companion’s orbit and the disk, we find\footnote{{The precession timescales are estimated by interpolating the contour scaling in Figure~3 of \citet{Batygin2012}, using representative points along the precession period curves. The code can be found} \href{https://github.com/wangxianyu7/Data_and_code/tree/main/ukb}{here}.} that the disk precession timescales are less than, or comparable to, $10\,\mathrm{Myr}$ for all misaligned systems (except HAT-P-7, WASP-94, WASP-100, and WASP-180, which have wide-separation and low-mass stellar companions, see Table~\ref{tab:binary}), implying that significant spin–orbit excitation by binary companions is possible \citep{Batygin2012,Lai2014}. This picture alone, however, does not explain why there is a relatively sharp change in the misaligned fraction at $\binarylambdacutvalue\,\mathrm{K}$ (\autoref{fig:kb}, panel a). 

We observe that, although the single-star obliquity transition is determined to be $\teff=\singlelambdacutvalue$~K based on the optimal decision boundary for KS tests, there are two systems cooler than this threshold that are misaligned (\autoref{fig:kb}, panel b): WASP-101 ($\teff = 6380 \pm 120$\,K, $\lambda = 34^\circ \pm 3$; \citealt{Zak2024A&A...686A.147Z}), and XO-4 ($\teff = 6397 \pm 70$\,K, $\lambda = 46\fdg7^{+6.1}_{-8.1}$; \citealt{Narita2010}). Our preceeding discussion suggests it to be plausible that these systems host as-yet-unidentified stellar companions. XO-4 has been observed with adaptive optics imaging \citep{Adams2013} with no companion detected, whereas WASP-101 has not yet been imaged at high angular resolution. Interestingly, the measured $\lambda$ values for these two systems lie very close to $35^\circ$. If we assume $\lambda\approx\psi$, these values are consistent with one of two peaks ($\sim35^\circ$ and $\sim115^\circ$) in the obliquity distribution predicted from from von Zeipel-Kozai-Lidov cycle simulations \citep{Fabrycky2007}. This coincidence suggests that these systems may indeed host undetected stellar companions; and targeted follow-up observations will be required to confirm or refute this possibility.

Either way, however, it remains puzzling how stellar companions could influence the rotation of the primary star, or tidal interactions with its planets, at the companion orbital separations relevant to our sample. Prior work shows that spin changes are most effective in close binaries through tidal interactions or companion-induced disk truncation/warping (i.e., at separations $\lesssim$ a few-tens of AU, \citealt{Meibom2007}). By contrast, most binaries in our sample have projected separations of order $10^2$-$10^3$ AU, in currently standard stellar-evolution recipes, systems at these separations are modeled as effectively isolated stars, with no meaningful cross-influence on interior structure or evolutionary tracks.

{Another possibility is that blended photometry/spectroscopy in binaries leads to systematic \teff\ biases (e.g., \citealt{Andersen1991,Torres2010,ElBadry2018,Furlan2020,Koenigsberger2025}), producing an apparent offset in the obliquity and rotation transitions relative to singles. Interestingly, we found that in our obliquity sample (consisting of hot Jupiters), the planet host is \emph{always} the more massive component of a binary (See \autoref{fig:teffgmag}).  This is consistent with the higher hot-Jupiter occurrence around higher-mass stars \citep{Johnson2010}, but it may also reflect detection bias, since transiting planets are more readily found around the brighter primary (e.g., \citealt{Wang2019TESS}). In such cases, it is tempting to suppose that a close, cooler companion with modest brightness contrast biases the composite spectrum/spectral energy distribution (SED) and drives the inferred \teff\ downward, artificially placing these systems into the ``cooler" regime. As shown in the Appendix~\ref{appendixA}, however, the host stars’ \teff\ values derived from multi-component SED modeling are consistent with the \teff\ adopted in this work from TEPCat, indicating that contamination of the primary’s SED by a stellar companion cannot explain the $\sim400~\mathrm{K}$ \teff\ shift. It is worth noting, however, that the SED fitting relies on spectroscopically derived \feh\ priors, which could still be biased by light contamination from stellar companions \citep{ElBadry2018}. Therefore, spatially resolved spectroscopic observations for the binaries listed in Table~\ref{tab:binary} are required to assess potential spectral contamination affecting the determination of the primaries’ \teff.}

\section{Implications of a higher Kraft break  }
\subsection{Few RMs above the single-star Kraft break}

A central open question in the study of stellar obliquity, as recently reviewed by \citet{Albrecht2022}, is whether spin-orbit misalignments are a byproduct of high-eccentricity migration, and therefore confined to hot-Jupiter-like systems (isolated close-in gas giants plausibly delivered by high-$e$ pathways), or instead reflect a universal process that is largely independent of planetary architecture.

Although, for practical reasons, most RM measurements have targeted hot Jupiters, NASA's \textit{K2} \citep{Howell2014} and \textit{TESS} \citep{Ricker2015} missions have provided suitable targets to extend obliquity measurements beyond hot Jupiters and have revealed intriguing trends. 

{The emerging picture has so far been that large spin-orbit misalignments concentrate in hot-Jupiter-like, isolated systems (including isolated sub-Saturns that may share high-$e$ delivery channels, \citealt{Yu2024}), whereas other categories of planetary systems are typically aligned (see Figure~\ref{fig:trends}), including compact {multi-planet} systems \citep{Albrecht2013, Wang2018a, Zhou2018,  WangX2022WASP148, Dai2023, Lubin2023, Radzom2024, Radzom2025}, warm Jupiters \citep{Wang2021,Rice2022WJs_Aligned, WangX2024}, eccentric Jupiters \citep{Espinoza2023,Mireles2025arXiv250922972M}, and very massive close-in companions with $m_p/M_\star\gtrsim 2\times10^{-3}$ (about $2\,M_{\rm Jup}$ for a solar-mass star, \citealt{Hebrard2011, Triaud2018, Albrecht2022, Zhou2019, Rusznak2025}), as well as brown dwarfs \citep[e.g.,][]{Giacalone2024,Ferreira2024,Brady2025,Carmichael2025arXiv250618971C,Doyle2025,Vowell2025arXiv251000105V,Zak2025arXiv250520516Z}. Taken at face value, this contrast may suggest that spin–orbit misalignment is not universally primordial across planetary architectures. }

Measuring the spin-orbit (mis)alignments of \emph{non-hot-Jupiter} planets around cool stars, {however}, provides little discriminating power between these hypotheses, because hot Jupiters around cool stars are also predominantly aligned. A decisive test instead requires obliquity measurements to be taken of non-hot-Jupiter planets around hot hosts. Thus, shifting the single-star Kraft break to $\sim 6500\,\mathrm{K}$ --- and, thereby, changing the boundary between which stars we classify as cool vs. hot --- changes the inference we are permitted to draw from the measurements we have in hand. At present there are only {five} such systems (CoRoT-3, \teff=$6558\pm44$ K, $\lambda=-37\fdg6^{+22.3}_{-10.0}$, \citealt{Triaud2009}; 
TOI-3362, \teff=$6800\pm100$~K, $\lambda=1\fdg2^{+2.8}_{-2.7}$
, \citealt{Espinoza2023}; 
TOI-558, \teff=$6466^{+95}_{-93}$~K, $\lambda=11\fdg7^{+7.5}_{-4.8}$
, \citealt{Espinoza2025};
TOI-778, \teff=$6643\pm150$~K, $\lambda=18\pm11$\degrees
, \citealt{Clark2023} and
WASP-120, \teff=$6450\pm120$~K, $\lambda=-2\pm4$\degrees
, \citealt{Zak2024A&A...686A.147Z}) with RM measurements around stars hotter than our single-star Kraft break, and only one of these (TOI-3362) lies beyond $1\sigma$ of the break (See \autoref{fig:trends}). With so few robust measurements around genuinely hot stars, the central question --- as to whether spin-orbit misalignment is intrinsically more common around hotter stars, regardless of planet type --- \emph{remains open}.

A further, albeit tentative, population-level line of evidence that hot stars may be \emph{primordially} more misaligned comes from the distributions of $\vsini$ \citep{Louden2021,Louden2024} and photometric modulation amplitudes \citep{Mazeh2015}: both are consistent with larger stellar tilts in hotter dwarfs, even for compact multi-planet systems. Moreover, among hot Jupiter hosts, misalignments appear more frequent around A-type than mid-F stars \citep{Albrecht2022}.

One of the strongest remaining arguments against primordial spin-orbit misalignment comes from sub-Saturns around cool stars (see the bottom panel of \autoref{fig:trends}
): single sub-Saturns around cool hosts are often misaligned \citep{Winnhatp11,Sanchis-Ojeda2011, Bourrier2022,Stefansson2022, Knudstrup2024}, whereas compact sub-Saturn multi-planet systems around similar cool hosts are typically well aligned \citep{Radzom2024, Radzom2025}. This contrast indicates that the misalignments of single sub-Saturns around cool stars are most likely dynamical, not primordial. That conclusion, however, should not be naively extrapolated to hot Jupiters, because mechanisms that readily excite the obliquity of a low-mass sub-Saturn do not necessarily operate with comparable efficiency for a massive hot Jupiter.

A concrete example of the perils of such extrapolation is secular resonance crossing \citep{Petrovich2020}, which {can drive sub-Saturns onto polar orbits} by transferring angular momentum deficit from an outer companion to an inner low-mass planet. Because the inner sub-Saturn carries less orbital angular momentum, its inclination can be strongly excited, whereas a hot Jupiter is much harder to tilt. This mechanism yields a testable prediction: around rapidly rotating hot stars, the stellar quadrupole can dominate over general-relativistic precession, detuning the resonance from exactly polar and driving a characteristic steady-state misalignment near $65^\circ$. Observationally, however, no sub-Saturn yet has an RM measurement above $\sim 6500\,\mathrm{K}$; recent attempts for TOI-1842 \citep{Wittenmyer2022, Hixenbaugh2023} and TOI-1135 \citep{Mallorquin2024, Dugan2025} target hosts that are hotter than the previously adopted $\sim6100\,\mathrm{K}$ boundary but lie just below the single-star Kraft break at $\sim 6500\,\mathrm{K}$, leaving the hot-star prediction (sub-polar clustering near $\sim65^\circ$) untested.

\subsection{Obliquity Damping Mechanisms}

Our newly attained precision in determining the location of the obliquity transition, complementing that also achieved for the location of the Kraft break, may shed light on the physical origin of the obliquity transition itself. So far, two tidal mechanisms have been proposed for obliquity damping that predict an obliquity transition --- the excitation of inertial waves which are then convectively dissipated \citep{Lai2012,saunders_efficient_2024}, vs. a net orbit-averaged tidal torque exerted by standing internal gravity waves (i.e., g-modes) which are resonance-locked with the orbital frequency \citep{zanazzi2024damping,Zanazzi2025} over the course of the long-term evolution of the \bv\ frequency. These proposals are associated with wave propagation and dissipation under different (although largely overlapping) regimes of stellar structure.

In particular, convectively-dissipated obliquity damping predicts an obliquity transition at main-sequence temperatures (and therefore, to leading order, stellar masses) at which stars lose their outer convective envelopes. At this point, magnetic rotational breaking also ceases. Therefore, the obliquity transition should coincide with the rotational Kraft break if this mechanism were solely responsible. By contrast, resonance-locking with g-modes results in an obliquity transition at a stellar mass limit, and therefore main-sequence effective temperature, primarily set by a transition in a specific feature in the long-term evolution of g-mode frequencies. This is most easily described in terms of evolutionary changes to the average \bv\ frequency,
\begin{equation}
	\left<N\right> = \int_\text{rad} \max(0, N) {\mathrm d r \over r}, \label{eq:bv}
\end{equation}
where $N$ is the \bv\ frequency. This integral runs from either the {center} of the star or the upper boundary of the convective core (if any), to the base of the outer convection zone.

\begin{figure}[htbp]
	\centering
	\includegraphics[width=.45\textwidth]{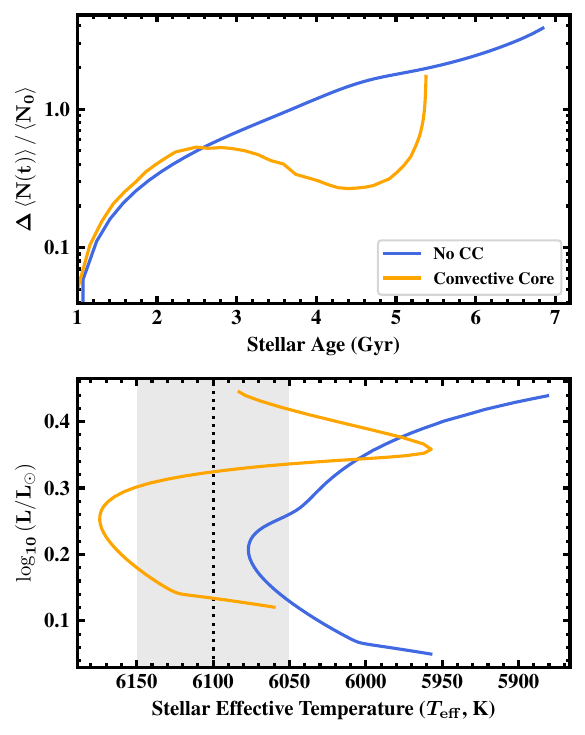}
	\caption{Evolution of solar-composition \texttt{MESA} stellar models on either side of the convective-core transition. In both panels, the blue curves show the time evolution of $1.11~M_\odot$ stellar models, which do not exhibit a long-lived convective core on the main sequence, while the orange curves show those of $1.15~M_\odot$ stellar models, where a convective core persists throughout the main sequence until core hydrogen exhaustion. The upper panel shows the time evolution of the average \bv\ frequency (or, equivalently, the g-mode undertone spacing), \autoref{eq:bv}, compared to a reference epoch of $t_0 = 1$ Gyr, as in \cite{zanazzi2024damping}. The lower panel shows how these stellar models evolve on the Hertzsprung-Russell diagram.}
	\label{fig:cc}
\end{figure}

We illustrate this feature of interest in the upper panel of \autoref{fig:cc}, using evolutionary stellar models on either side of this transition. These models\footnote{{Our MESA inlist files, and a shell script to generate the evolutionary tracks used in this work, have been made available as a deposit on \dataset[Zenodo]{https://doi.org/10.5281/zenodo.17470968}.}} were generated with \texttt{MESA} r12778 \citep{mesa_paper_1,mesa_paper_2,mesa_paper_3,mesa_paper_4,mesa_paper_5}, using solar-calibrated values of the initial helium abundance and mixing length parameter. Both account for elemental diffusion, and include a small amount of convective overmixing, with $f_\text{ov} = 0.05$. Over the course of main-sequence evolution, \citet{zanazzi2024damping} and \citet{Zanazzi2025} show that the averaged \bv\ frequencies of cool stars, and therefore all of their g-mode frequencies, increase significantly, gradually, and monotonically over the course of stellar evolution compared to their values $\left<N_0\right> = \left<N(t_0)\right>$ at some reference epoch $t_0$ of planet formation. By contrast, stars that are hotward of the Kraft break have g-modes whose frequencies actually decrease over substantial portions of their main-sequence lifetime, and rapidly increase near main-sequence turnoff. As a result, g-modes in these cool stars are capable of maintaining resonance locks (and, correspondingly, sustaining obliquity-damping torques) during the gradual inward migration of their planets, while those in stars hotward of the Kraft break are not. Thus, were resonance-locking tidal torques to be the dominant mechanism of obliquity damping, an obliquity transition should occur at the same temperature at which this transition in the behaviour of the g-mode frequencies also occurs.

We now make the observation that this transition is distinct from the Kraft break. Both sets of evolutionary models shown in the lower panel of \autoref{fig:cc} have very similar stellar masses ($1.11$ vs $1.15~M_\odot$), and very similarly sized convective envelopes, with quite significant depths (reaching to $0.76$ and $0.78~R_\star$, respectively). Rather, the cause of this transition is that the lower-mass, cooler, set of evolutionary models is not massive enough to possess a convective core, while the higher-mass set of evolutionary models sustains a convective core throughout its main-sequence lifetime.  As in \citet{zanazzi2024damping} and \citet{Zanazzi2025}, our MESA modelling indicates that this transition occurs at a main-sequence effective temperature of roughly $6100 \pm 100$ K, as can be seen in the lower panel of \autoref{fig:cc}. In general, the location of this stellar-mass boundary is also sensitive to other properties of the star, and to the physics adopted in stellar modelling, in particular the amount of overshooting/convective boundary mixing. However, it is known to be substantially cooler (being at a lower stellar mass) than the rotational Kraft break --- e.g. \citet{Silva2013,bellinger_inversion_2019,buchele_inversion_2025}. So too, therefore, is the obliquity transition predicted by the resonance-locking mechanism alone.

In practice, these two mechanisms are not likely to be mutually exclusive. Both may operate in stars which are cool enough to have convective envelopes and no convective cores ($\teff < 6100$~K), whereas neither mechanism is expected to operate in stars with convective cores and no convective envelopes ($\teff>6500$~K). The two misaligned single-star systems discussed in Section~4, with $6100\!<\!\teff\!<\!6500$~K, then naturally fall into a regime where only inertial-wave tides operate; the damping is consequently weaker than in cooler stars where both channels are available.  Nevertheless, the fact that the Kraft rotation break coincides with the single-star obliquity transition does at least strongly (if circumstantially) suggest that the loss of convective dissipation at higher stellar masses, rather than of resonance locking, is responsible for the obliquity transition.

\section{Conclusion}

It is a common working assumption that the stellar-obliquity transition coincides with the rotational Kraft break, motivating the view that obliquity damping in cool stars drives alignment \citep{Winn2010, Albrecht2012, WinnFabrycky2015, Albrecht2022}. However, it has been noted \citep{Beyer2024}, and our analysis confirms, that the commonly quoted obliquity break near $\teff=6100~\mathrm{K}$ is several hundred kelvin cooler than the classical rotational break around $6500~\mathrm{K}$, posing a fundamental inconsistency.

\vspace{1cm}
In this work, we have found: 
\begin{enumerate}
\item \textbf{A unified single-star break at $\sim$ 6500~K.} Restricting to \emph{single stars} (Panel b of \autoref{fig:kb}), the obliquity transition occurs at $\singlelambdacut~\mathrm{K}$, in excellent agreement with the single-star rotational Kraft break ($\singlevsinicut~\mathrm{K}$, panel c of \autoref{fig:kb}).

\item \textbf{A cooler break arises in multiples.} The $\sim6100~\mathrm{K}$ obliquity break appears when \emph{binaries/multiples} are included (panels a and d of \autoref{fig:kb}), although the reason is ambiguous (See \autoref{sec:binaries} for a detailed discussion).
\end{enumerate}

\medskip
This result is consequential for our understanding of some key issues: 

\begin{enumerate}
\item \textbf{The origin of spin-orbit misalignment (Hot Jupiter architecture vs.\ hot hosts).} With the single-star obliquity break revised upward to $\sim 6500\,\mathrm{K}$, non-hot-Jupiter RM measurements around \emph{truly} hot hosts $\left(\teff \gtrsim 6500\,\mathrm{K}\right)$ become extremely sparse (see \autoref{fig:trends}), leaving open whether large misalignments are unique to Hot Jupiter architectures or broadly common around hot stars.

\item \textbf{The evolution of spin-orbit misalignment (damping mechanism: convective vs.\ radiative).} The near-perfect coincidence between the single-star obliquity transition and the rotational Kraft break points to a common physical cause: as outer convective envelopes thin toward higher $\teff$, inertial-wave tidal obliquity damping and magnetic braking both weaken, naturally producing simultaneous transitions. The observed coincidence disfavors resonance-locking as the main source of obliquity damping, because it alone would predict an obliquity transition tied to the \emph{convective-core} onset (via g-mode frequency evolution) at $\teff=6100\pm100~\mathrm{K}$, distinctly cooler than the $\sim6500~\mathrm{K}$ single-star rotation break set by the thinning/vanishing outer convective envelope. We do not, however, claim that these tidal mechanisms must act mutually exclusively; resonance locking may still operate alongside convective-envelope tides in appropriate regimes.
\end{enumerate}

\clearpage
{We thank the anonymous reviewer for their constructive feedback, which has greatly improved the quality of this manuscript.} We thank Jie Yu and Zhao Guo for bringing the \citet{Beyer2024} paper to our attention. We are grateful for helpful discussions with {J. J. Zanazzi, Eugene Chiang, Yubo Su, Yanqin Wu, Janosz Dewberry, Fred C. Adams,  Simon H. Albrecht, Joshua N. Winn, Fei Dai}, Constantine Deliyannis, Caty Pilachowski, Kareem El-Badry, Cheyanne Shariat, Gang Li, Meng Sun, Cristobal Petrovich, Brandon Radzom, Gongjie Li, Malena Rice,  Samir Salim, Zachary Maas, and Timothy Bedding. This work was supported in part by the NASA Exoplanets Research Program  NNH23ZDA001N-XRP (Grant No. 80NSSC24K0153), the NASA TESS General Investigator Program, Cycle~7, NNNH23ZDA001N-TESS (Grant No. 80NSSC25K7912), and the Heising-Simons Foundation (Grant \#2023-4050). {S.W. gratefully acknowledges support from the John and A-Lan Reynolds Faculty Research Fund, which enabled participation in the International Conference on Exoplanets and Planet Formation (EPF). This support was vital for presenting this result and discussing the puzzling observation with the world's leading researchers in the field.} Additionally, X.Y.W. acknowledges support from the Sullivan Prize Fellowship. J.M.J.O. acknowledges support from NASA through the NASA Hubble Fellowship grant HST-HF2-51517.001, awarded by STScI, which is operated by the Association of Universities for Research in Astronomy, Incorporated, under NASA contract NAS5-26555. This research was supported in part by Lilly Endowment, Inc., through its support for the Indiana University Pervasive Technology Institute.


\vspace{5mm}
\facilities{WIYN/NEID, PFS/Magellan}

\software{\texttt{numpy} \citep{oliphant2006guide, walt2011numpy, harris2020array}, \texttt{matplotlib} \citep{hunter2007matplotlib}, \texttt{pandas} \citep{mckinney2010data}, \texttt{scipy} \citep{virtanen2020scipy}, \texttt{MESA} \citep{mesa_paper_1,mesa_paper_2,mesa_paper_3,mesa_paper_4,mesa_paper_5}}

\clearpage
\appendix

\begin{deluxetable}{lcccccccccccccc}
\tablecaption{\textbf{Misaligned Hot Jupiters in Binaries or Multi-star systems} \label{tab:binary}}
\tabletypesize{\scriptsize}
\tablehead{
\colhead{} & \multicolumn{3}{c}{\teff\, adopted by TEPCat}  & \colhead{}& \colhead{}& \colhead{} & 
\multicolumn{5}{c}{Stellar companions' properties} &\multicolumn{2}{c}{Primary \teff\, from MS$^a$} \\
\cline{2-4}\cline{8-11}\cline{13-14}
\colhead{Primary} & \colhead{\teff} & \colhead{Method$^{a}$} & \colhead{Ref.}  &\colhead{$\lambda$}&\colhead{Ref.}& \colhead{Companion} & 
\colhead{Sep} & \colhead{Sep} &  \colhead{Mass} & \colhead{Ref.} & \colhead{$\Delta$Mag$^{f}$} & \colhead{\teff}& \colhead{Ref.}\\
\colhead{} & \colhead{(K)} &  \colhead{} &\colhead{(\degrees)} &\colhead{}&\colhead{} & \colhead{} & 
\colhead{(\arcsec)} & \colhead{(AU)} &  \colhead{(\msun)} & \colhead{} & \colhead{}& \colhead{(K)}& \colhead{}&\colhead{} 
}
\startdata
\multicolumn{2}{l}{\textbf{Binaries and multiples}} \smallskip\\
HAT-P-7 A   & $6310\pm15$  & AM     & 1 & $155\pm37$  & 2 & B & 3.86 & $1504^{+1377}_{-521}$$^{c}$  & $0.38^{+0.12}_{-0.11}$   & 3 & 8.2 (G) & $6360\pm120$ & 4 \\
HAT-P-14 A  & $6600\pm90$  & SF     & 5 & $189.1\pm5.1$ & 6 & B & 0.87 & $180\pm10$      & $0.20\pm0.04$   & 7 & 5.8 (K$_s$) & 6540$\pm$120 & 4 \\
HAT-P-30 A  & $6338\pm42$  & SF     & 8 & $73.5\pm9.0$ & 9 & B & 3.83 & $882^{+430}_{-201}$$^{c}$     & $0.58^{+0.10}_{-0.11}$   & 3 & 4.9 (G) & $6190\pm110$ & 4 \\
HAT-P-32 A  & $6269\pm64$  & MS     & 10 & $85\pm1.5$   & 2 & B & 2.94 & $831\pm15$                    & $0.4243\pm0.0085$        & 7 & 6.1 (G) & $6269\pm64$ & 10 \\
KELT-4 A    & $6206\pm75$  & MS     & 11 & $80^{+25}_{-22}$ & 12 & B/C$^{b}$ & 1.5  & $328\pm16$              & $0.65\pm0.10$             & 11 & 1.4 (K$_p$)  & $6206\pm75$ & 11 \\
KELT-18 A    & $6670\pm120$  & MS     & 13 & $-94.8\pm0.7$ & 14 & B & 3.43  & 1100   &$0.653^{+0.037}_{-0.04}$$^{d}$     & 13 & 5.4 (K$_p$)  & $6670\pm120$ & 13 \\
KELT-23 A   & $5899\pm59$  & MS     & 15 & $180.4^{+4.9}_{-4.7}$ & 16 & B & 4.5 & 570 & 0.25 & 15 & 5.4 (G) & $5899\pm59$  & 15 \\
\multirow{2}{*}{WASP-12 A}  & \multirow{2}{*}{$6265\pm50$} & \multirow{2}{*}{SF} & \multirow{2}{*}{17} & \multirow{2}{*}{$59^{+15}_{-20}$} & \multirow{2}{*}{2} & B & 1.06 & $462\pm39$ & $0.554\pm0.020$ & 7 & 3.3 (K$_s$) & \multirow{2}{*}{$6140^{+130}_{-120}$} & \multirow{2}{*}{4} \\
   &  &  &  &  &  & C & 1.07 & $466\pm40$ & $0.571\pm0.019$ & 7 & 3.2 (K$_s$) &  & \\
WASP-76 A   & $6329\pm65$  & SF     & 18 & $-37^{+13}_{-21}$ & 18 & B & 0.44 & $53.0\pm8.8$ & $0.712\pm0.042$ & 19 & 2.7 (K$_s$) & $6366^{+92}_{-90}$ & 20 \\
WASP-94 A   & $6170\pm80$  & SS     & 21 & $151^{+16}_{-23}$ & 22 & B & 16 & 2700 & $1.24\pm0.09$ & 21 & 0.4 (G) & $6170\pm80$ & 21 \\
WASP-100 A   & $6940\pm120$  & SS     & 23 & $79^{+19}_{-10}$ & 24 & B & 3.96 & $1547^{+895}_{-403}$ & $0.45\pm0.11$ & 3 & 6.9 (G) & 6560$\pm$180 & 4 \\
WASP-136 A  & $6250\pm100$ & SF     & 25 & $-38^{+16}_{-18}$ & 12 & B & 5.07 & 1364 & --$^{e}$ & 26 & 9.8 (G)& $6360^{+150}_{-140}$ & 4 \\
WASP-180 A  & $6500\pm150$ & SF     & 27 & $-157\pm2$ & 27 & B & 5 & 1200 & 1 & 27 & 0.9 (G)& $6150\pm120$ & 4 \\
\multicolumn{2}{l}{\textbf{Binary candidates}} \smallskip\\
HAT-P-50    & $6280\pm49$  & SF     & 28 & $41^{+10}_{-9}$ & 22 & B & -- & -- & -- & -- & --& -- & -- \\
K2-237      & $6099\pm110$ & SF     & 29 & $91\pm7$ & 24 & B & -- & -- & -- & -- & --& $6110^{+120}_{-130}$ & 4 \\
\enddata
\tablenotetext{}{$^{*}$
1~\citet{Benomar2014},
2~\citet{Albrecht2012},
3~\citet{Rice2024},
4~This work,
5~\citet{Torres2010HATP14},
6~\citet{Winn2011},
7~\citet{Ngo2015},
8~\citet{Mortier2013},
9~\citet{Johnson2011},
10~\citet{Zhao2014},
11~\citet{Eastman2016},
12~\citet{Knudstrup2024},
13~\citet{McLeod2017},
14~\citet{Rubenzahl2024},
15~\citet{Johns2019},
16~\citet{Giacalone2025},
17~\citet{Leonardi2024},
18~\citet{Bourrier2024},
19~\citet{Ngo2016},
20~\citet{Fu2021},
21~\citet{Stassun2017},
22~\citet{NeveuVanMalle2014},
23~\citet{Jansen2020},
24~\citet{Addison2018},
25~\citet{Lam2017},
26~\citet{Badry2021},
27~\citet{Temple2019},
28~\citet{Hartman2015},
29~\citet{Smith2019}.
\\$^{a}$
AM: asteroseismic modeling, SF: spectra fitting, MS: multi-component SED fitting, SS:  single-component SED fitting.
\\
$^{b}$
Given that \citet{Eastman2016} estimated KELT-4 B and C to be twin K-type stars, we assume they share identical properties.\\
$^{c}$
This value was derived using \texttt{lofti$\_$gaia} \citep{Pearce2020}, incorporating the companion mass reported by \citet{Rice2024} and astrometric measurements (relative RA and DEC, relative proper motions, and \textit{Gaia} parallax) from the \textit{Gaia} DR3 catalog \citep{GaiaCollaboration2023}.\\
$^{d}$
From our multi-component SED modeling.\\
$^{e}$
Only the \textit{Gaia} $G$ magnitude is available, so the stellar mass cannot be derived through isochrone fitting.
\\
$^{f}$
G: $Gaia$ G band, K$_p$ or K$_s$: Keck NIRC2 K bands.\\
}
\end{deluxetable}

\section{{Multi-component SED modeling for systems with stellar companions}\label{appendixA}}

{Based on the misalignment criteria ($|\lambda| - 2\sigma_{\lambda} > 0$ and $|\lambda| > 10\degrees$), we identified 15 misaligned hot-Jupiter systems that reside in (candidate) binary or multiple-star systems: HAT-P-7, HAT-P-14, HAT-P-30, HAT-P-32, HAT-P-50, K2-237, KELT-4, KELT-18, KELT-23, WASP-12, WASP-76, WASP-94, WASP-100, WASP-100, and WASP-136. The properties of these systems are listed in Table~\ref{tab:binary}.}

{The separation between WASP-94 A and WASP-94 B is 15$\arcsec$, which is well resolved by $Gaia$ DR3 \citep{GaiaCollaboration2023}, 2MASS \citep{Cutri2003}, and WISE \citep{Cutri2014}, and thus light from the stellar companion does not affect the \teff\ determination of the primary. For HAT-P-32 A, KELT-4 A, KELT-18 A, and WASP-76 A, the \teff\, values adopted by TEPCat are derived from multi-component SED fittings. HAT-P-50 is a binary candidate with a \textit{Gaia} DR3 RUWE of 1.4, but no stellar companion has been confirmed, so we cannot estimate the level of possible light contamination from a companion. Therefore, these systems were excluded from our SED analysis.}

{For HAT-P-7, HAT-P-14, HAT-P-30, WASP-12, WASP-100, and WASP-136, we performed multi-component SED modeling using \texttt{EXOFASTv2} \citep{Eastman2017,Eastman2019} to derived non-contamination SED-based \teff\, for primary stars. We adopted broadband photometry including \textit{Gaia} $G$, $G_{\rm bp}$, $G_{\rm rp}$, 2MASS $JHK$, and WISE 1–3 for primary stars. For the stellar companions, we included broad photometry from \textit{Gaia} DR3 together with direct imaging measurements \citep{Ngo2015,McLeod2017}. Gaussian priors on \feh\ were derived from the SWEET-Cat catalog \citep{Santos2013sweetcat,Sousa2021}, which provides spectroscopic stellar parameters derived from a uniform and homogeneous analysis of the spectra. We assumed that the primary and companion stars share the same initial \feh, $V$-band extinction, distance, and age, and adopted the $V$-band extinction value from \citet{Schlafly2011} as the upper limit on extinction. The Gaia DR3 parallax corrected for the zero-point offset, along with its uncertainty \citep{GaiaCollaboration2023, Lindegren2021}, was adopted as a Gaussian prior on the parallax.}

{WASP-180 A and B are resolved in \textit{Gaia} DR3, 2MASS, and WISE, but no \teff\, from SED solution is available. We therefore performed single-component SED fit using the aforementioned broad photometry. For KELT-18, although \citet{McLeod2017} performed multi-component SED modeling, they did not report the stellar companion’s mass. We therefore carried out our own multi-component SED modeling to derive it. K2-237 lies in a crowded stellar field, making it impossible to identify bound companions. Therefore, we treated it as a single star and performed single-star SED modeling using \textit{Gaia} $G$, $G_{\rm bp}$, $G_{\rm rp}$, and 2MASS $JHK$, excluding WISE 1–3 photometry due to potential light contamination from nearby stars.}

{We adopted standard convergence criteria (independent draws $T_{z}$ $>$ 1000 and Gelman–Rubin statistic $\Hat{R}<$ 1.01; \citealt{Eastman2019}). Our analysis, summarized in Table~\ref{tab:binary}, shows that, for primary stars, the \teff\ values from the uncontaminated SED modeling remain consistent, within $3\sigma$, with the \teff\, values adopted in this work from TEPCat.}

\bibliography{main.bib}

@ARTICLE{Ngo2016,
       author = {{Ngo}, Henry and {Knutson}, Heather A. and {Hinkley}, Sasha and {Bryan}, Marta and {Crepp}, Justin R. and {Batygin}, Konstantin and {Crossfield}, Ian and {Hansen}, Brad and {Howard}, Andrew W. and {Johnson}, John A. and {Mawet}, Dimitri and {Morton}, Timothy D. and {Muirhead}, Philip S. and {Wang}, Ji},
        title = "{Friends of Hot Jupiters. IV. Stellar Companions Beyond 50 au Might Facilitate Giant Planet Formation, but Most are Unlikely to Cause Kozai-Lidov Migration}",
      journal = {\apj},
         year = 2016,
        month = aug,
       volume = {827},
       number = {1},
          eid = {8},
        pages = {8},
          doi = {10.3847/0004-637X/827/1/8},
archivePrefix = {arXiv},
       eprint = {1606.07102},
 primaryClass = {astro-ph.EP},
       adsurl = {https://ui.adsabs.harvard.edu/abs/2016ApJ...827....8N}
}

@ARTICLE{Lam2017,
       author = {{Lam}, K.~W.~F. and {Faedi}, F. and {Brown}, D.~J.~A. and {Anderson}, D.~R. and {Delrez}, L. and {Gillon}, M. and {H{\'e}brard}, G. and {Lendl}, M. and {Mancini}, L. and {Southworth}, J. and {Smalley}, B. and {Triaud}, A.~H.~M. and {Turner}, O.~D. and {Hay}, K.~L. and {Armstrong}, D.~J. and {Barros}, S.~C.~C. and {Bonomo}, A.~S. and {Bouchy}, F. and {Boumis}, P. and {Collier Cameron}, A. and {Doyle}, A.~P. and {Hellier}, C. and {Henning}, T. and {Jehin}, E. and {King}, G. and {Kirk}, J. and {Louden}, T. and {Maxted}, P.~F.~L. and {McCormac}, J.~J. and {Osborn}, H.~P. and {Palle}, E. and {Pepe}, F. and {Pollacco}, D. and {Prieto-Arranz}, J. and {Queloz}, D. and {Rey}, J. and {S{\'e}gransan}, D. and {Udry}, S. and {Walker}, S. and {West}, R.~G. and {Wheatley}, P.~J.},
        title = "{From dense hot Jupiter to low-density Neptune: The discovery of WASP-127b, WASP-136b, and WASP-138b}",
      journal = {\aap},
         year = 2017,
        month = mar,
       volume = {599},
          eid = {A3},
        pages = {A3},
          doi = {10.1051/0004-6361/201629403},
archivePrefix = {arXiv},
       eprint = {1607.07859},
 primaryClass = {astro-ph.EP},
       adsurl = {https://ui.adsabs.harvard.edu/abs/2017A&A...599A...3L}
}

@ARTICLE{Bourrier2024,
       author = {{Bourrier}, V. and {Delisle}, J.-B. and {Lovis}, C. and {Cegla}, H.~M. and {Cretignier}, M. and {Allart}, R. and {Al Moulla}, K. and {Tavella}, S. and {Attia}, M. and {Mounzer}, D. and {Vaulato}, V. and {Steiner}, M. and {Vrignaud}, T. and {Mercier}, S. and {Dumusque}, X. and {Ehrenreich}, D. and {Seidel}, J.~V. and {Wyttenbach}, A. and {Dethier}, W. and {Pepe}, F.},
        title = "{The ANTARESS workflow: I. Optimal extraction of spatially resolved stellar spectra with high-resolution transit spectroscopy}",
      journal = {\aap},
         year = 2024,
        month = nov,
       volume = {691},
          eid = {A113},
        pages = {A113},
          doi = {10.1051/0004-6361/202449203},
archivePrefix = {arXiv},
       eprint = {2407.19012},
 primaryClass = {astro-ph.EP},
       adsurl = {https://ui.adsabs.harvard.edu/abs/2024A&A...691A.113B}
}

@ARTICLE{Giacalone2025,
       author = {{Giacalone}, Steven and {Howard}, Andrew W. and {Rubenzahl}, Ryan A. and {Dai}, Fei and {Handley}, Luke B. and {Isaacson}, Howard and {Halverson}, Samuel and {Brodheim}, Max and {Brown}, Matt and {Carmichael}, Theron W. and {Deich}, William and {Fulton}, Benjamin J. and {Gibson}, Steven R. and {Hill}, Grant M. and {Holden}, Bradford and {Householder}, Aaron and {Laher}, Russ R. and {Lanclos}, Kyle and {Payne}, Joel and {Petigura}, Erik A. and {Roy}, Arpita and {Schwab}, Christian and {Sirk}, Martin M. and {Walawender}, Josh},
        title = "{A Hot Jupiter with a Retrograde Orbit Around A Sun-like Star and a Toy Model of Hot Jupiters in Wide Binary Star Systems}",
      journal = {\pasp},
         year = 2025,
        month = jul,
       volume = {137},
       number = {7},
          eid = {074401},
        pages = {074401},
          doi = {10.1088/1538-3873/adecc2},
archivePrefix = {arXiv},
       eprint = {2507.02667},
 primaryClass = {astro-ph.EP},
       adsurl = {https://ui.adsabs.harvard.edu/abs/2025PASP..137g4401G}
}

@ARTICLE{Pearce2020,
       author = {{Pearce}, Logan A. and {Kraus}, Adam L. and {Dupuy}, Trent J. and {Mann}, Andrew W. and {Newton}, Elisabeth R. and {Tofflemire}, Benjamin M. and {Vanderburg}, Andrew},
        title = "{Orbital Parameter Determination for Wide Stellar Binary Systems in the Age of Gaia}",
      journal = {\apj},
         year = 2020,
        month = may,
       volume = {894},
       number = {2},
          eid = {115},
        pages = {115},
          doi = {10.3847/1538-4357/ab8389},
archivePrefix = {arXiv},
       eprint = {2003.11106},
 primaryClass = {astro-ph.SR},
       adsurl = {https://ui.adsabs.harvard.edu/abs/2020ApJ...894..115P}
}

@ARTICLE{Hobbs1986,
       author = {{Hobbs}, L.~M. and {Pilachowski}, Catherine},
        title = "{Lithium in NGC 752}",
      journal = {\apjl},
         year = 1986,
        month = oct,
       volume = {309},
        pages = {L17},
          doi = {10.1086/184752},
       adsurl = {https://ui.adsabs.harvard.edu/abs/1986ApJ...309L..17H}
}

@ARTICLE{Feinstein2020,
       author = {{Feinstein}, Adina D. and {Montet}, Benjamin T. and {Ansdell}, Megan and {Nord}, Brian and {Bean}, Jacob L. and {G{\"u}nther}, Maximilian N. and {Gully-Santiago}, Michael A. and {Schlieder}, Joshua E.},
        title = "{Flare Statistics for Young Stars from a Convolutional Neural Network Analysis of TESS Data}",
      journal = {\aj},
         year = 2020,
        month = nov,
       volume = {160},
       number = {5},
          eid = {219},
        pages = {219},
          doi = {10.3847/1538-3881/abac0a},
archivePrefix = {arXiv},
       eprint = {2005.07710},
 primaryClass = {astro-ph.SR},
       adsurl = {https://ui.adsabs.harvard.edu/abs/2020AJ....160..219F}
}

@ARTICLE{Mallorquin2024,
       author = {{Mallorqu{\'\i}n}, M. and {Lodieu}, N. and {B{\'e}jar}, V.~J.~S. and {Zapatero Osorio}, M.~R. and {Sanz-Forcada}, J. and {Alarcon}, M.~R. and {Tabernero}, H.~M. and {Nagel}, E. and {Collins}, K.~A. and {Ciardi}, D.~R. and {Serra-Ricart}, M. and {Orell-Miquel}, J. and {Barkaoui}, K. and {Burdanov}, A. and {de Wit}, J. and {Everett}, M.~E. and {Gillon}, M. and {Jensen}, E.~L.~N. and {Murphy}, L.~G. and {Reed}, P.~A. and {Safonov}, B. and {Strakhov}, I.~A. and {Ziegler}, C.},
        title = "{TOI-1135 b: A young hot Saturn-size planet orbiting a solar-type star}",
      journal = {\aap},
         year = 2024,
        month = may,
       volume = {685},
          eid = {A90},
        pages = {A90},
          doi = {10.1051/0004-6361/202349016},
archivePrefix = {arXiv},
       eprint = {2402.17448},
 primaryClass = {astro-ph.EP},
       adsurl = {https://ui.adsabs.harvard.edu/abs/2024A&A...685A..90M}
}

@ARTICLE{Struve1931,
       author = {{Struve}, O. and {Elvey}, C.~T.},
        title = "{Algol and stellar rotation}",
      journal = {\mnras},
         year = 1931,
        month = apr,
       volume = {91},
        pages = {663},
          doi = {10.1093/mnras/91.6.663},
       adsurl = {https://ui.adsabs.harvard.edu/abs/1931MNRAS..91..663S}
}

@ARTICLE{Westgate1934,
       author = {{Westgate}, Christine},
        title = "{A Statistical Study of Rotational Broadening in 112 Stars of Class F}",
      journal = {\apj},
         year = 1934,
        month = apr,
       volume = {79},
        pages = {357},
          doi = {10.1086/143542},
       adsurl = {https://ui.adsabs.harvard.edu/abs/1934ApJ....79..357W}
}

@ARTICLE{Wittenmyer2022,
       author = {{Wittenmyer}, Robert A. and {Clark}, Jake T. and {Trifonov}, Trifon and {Addison}, Brett C. and {Wright}, Duncan J. and {Stassun}, Keivan G. and {Horner}, Jonathan and {Lowson}, Nataliea and {Kielkopf}, John and {Kane}, Stephen R. and {Plavchan}, Peter and {Shporer}, Avi and {Zhang}, Hui and {Bowler}, Brendan P. and {Mengel}, Matthew W. and {Okumura}, Jack and {Rabus}, Markus and {Johnson}, Marshall C. and {Harbeck}, Daniel and {Tronsgaard}, Ren{\'e} and {Buchhave}, Lars A. and {Collins}, Karen A. and {Collins}, Kevin I. and {Gan}, Tianjun and {Jensen}, Eric L.~N. and {Howell}, Steve B. and {Furlan}, E. and {Gnilka}, Crystal L. and {Lester}, Kathryn V. and {Matson}, Rachel A. and {Scott}, Nicholas J. and {Ricker}, George R. and {Vanderspek}, Roland and {Latham}, David W. and {Seager}, S. and {Winn}, Joshua N. and {Jenkins}, Jon M. and {Rudat}, Alexander and {Quintana}, Elisa V. and {Rodriguez}, David R. and {Caldwell}, Douglas A. and {Quinn}, Samuel N. and {Essack}, Zahra and {Bouma}, Luke G.},
        title = "{TOI-1842b: A Transiting Warm Saturn Undergoing Reinflation around an Evolving Subgiant}",
      journal = {\aj},
         year = 2022,
        month = feb,
       volume = {163},
       number = {2},
          eid = {82},
        pages = {82},
          doi = {10.3847/1538-3881/ac3f39},
archivePrefix = {arXiv},
       eprint = {2112.00198},
 primaryClass = {astro-ph.EP},
       adsurl = {https://ui.adsabs.harvard.edu/abs/2022AJ....163...82W}
}

@ARTICLE{Hebrard2011,
       author = {{H{\'e}brard}, G. and {Ehrenreich}, D. and {Bouchy}, F. and {Delfosse}, X. and {Moutou}, C. and {Arnold}, L. and {Boisse}, I. and {Bonfils}, X. and {D{\'\i}az}, R.~F. and {Eggenberger}, A. and {Forveille}, T. and {Lagrange}, A. -M. and {Lovis}, C. and {Pepe}, F. and {Perrier}, C. and {Queloz}, D. and {Santerne}, A. and {Santos}, N.~C. and {S{\'e}gransan}, D. and {Udry}, S. and {Vidal-Madjar}, A.},
        title = "{The retrograde orbit of the HAT-P-6b exoplanet}",
      journal = {\aap},
         year = 2011,
        month = mar,
       volume = {527},
          eid = {L11},
        pages = {L11},
          doi = {10.1051/0004-6361/201016331},
archivePrefix = {arXiv},
       eprint = {1101.5009},
 primaryClass = {astro-ph.EP},
       adsurl = {https://ui.adsabs.harvard.edu/abs/2011A&A...527L..11H}
}

@ARTICLE{Winn2009,
       author = {{Winn}, Joshua N. and {Howard}, Andrew W. and {Johnson}, John Asher and {Marcy}, Geoffrey W. and {Gazak}, J. Zachary and {Starkey}, Donn and {Ford}, Eric B. and {Col{\'o}n}, Knicole D. and {Reyes}, Francisco and {Nortmann}, Lisa and {Dreizler}, Stefan and {Odewahn}, Stephen and {Welsh}, William F. and {Kadakia}, Shimonee and {Vanderbei}, Robert J. and {Adams}, Elisabeth R. and {Lockhart}, Matthew and {Crossfield}, Ian J. and {Valenti}, Jeff A. and {Dantowitz}, Ronald and {Carter}, Joshua A.},
        title = "{The Transit Ingress and the Tilted Orbit of the Extraordinarily Eccentric Exoplanet HD 80606b}",
      journal = {\apj},
         year = 2009,
        month = oct,
       volume = {703},
       number = {2},
        pages = {2091-2100},
          doi = {10.1088/0004-637X/703/2/2091},
archivePrefix = {arXiv},
       eprint = {0907.5205},
 primaryClass = {astro-ph.EP},
       adsurl = {https://ui.adsabs.harvard.edu/abs/2009ApJ...703.2091W}
}

@ARTICLE{Winnhatp11,
       author = {{Winn}, Joshua N. and {Johnson}, John Asher and {Howard}, Andrew W. and {Marcy}, Geoffrey W. and {Isaacson}, Howard and {Shporer}, Avi and {Bakos}, G{\'a}sp{\'a}r {\'A}. and {Hartman}, Joel D. and {Albrecht}, Simon},
        title = "{The Oblique Orbit of the Super-Neptune HAT-P-11b}",
      journal = {\apjl},
         year = 2010,
        month = nov,
       volume = {723},
       number = {2},
        pages = {L223-L227},
          doi = {10.1088/2041-8205/723/2/L223},
archivePrefix = {arXiv},
       eprint = {1009.5671},
 primaryClass = {astro-ph.EP},
       adsurl = {https://ui.adsabs.harvard.edu/abs/2010ApJ...723L.223W}
}

@ARTICLE{Sanchis-Ojeda2011,
       author = {{Sanchis-Ojeda}, Roberto and {Winn}, Joshua N.},
        title = "{Starspots, Spin-Orbit Misalignment, and Active Latitudes in the HAT-P-11 Exoplanetary System}",
      journal = {\apj},
         year = 2011,
        month = dec,
       volume = {743},
       number = {1},
          eid = {61},
        pages = {61},
          doi = {10.1088/0004-637X/743/1/61},
archivePrefix = {arXiv},
       eprint = {1107.2920},
 primaryClass = {astro-ph.EP},
       adsurl = {https://ui.adsabs.harvard.edu/abs/2011ApJ...743...61S}
}

@ARTICLE{Mazeh2015,
       author = {{Mazeh}, Tsevi and {Perets}, Hagai B. and {McQuillan}, Amy and {Goldstein}, Eyal S.},
        title = "{Photometric Amplitude Distribution of Stellar Rotation of KOIs{\textemdash}Indication for Spin-Orbit Alignment of Cool Stars and High Obliquity for Hot Stars}",
      journal = {\apj},
         year = 2015,
        month = mar,
       volume = {801},
       number = {1},
          eid = {3},
        pages = {3},
          doi = {10.1088/0004-637X/801/1/3},
archivePrefix = {arXiv},
       eprint = {1501.01288},
 primaryClass = {astro-ph.EP},
       adsurl = {https://ui.adsabs.harvard.edu/abs/2015ApJ...801....3M}
}

@ARTICLE{Louden2024,
       author = {{Louden}, Emma M. and {Wang}, Songhu and {Winn}, Joshua N. and {Petigura}, Erik A. and {Isaacson}, Howard and {Handley}, Luke and {Yee}, Samuel W. and {Beard}, Corey and {Murphy}, Joseph M. Akana and {Laughlin}, Gregory},
        title = "{A Larger Sample Confirms Small Planets around Hot Stars Are Misaligned}",
      journal = {\apjl},
         year = 2024,
        month = jun,
       volume = {968},
       number = {1},
          eid = {L2},
        pages = {L2},
          doi = {10.3847/2041-8213/ad4b1b},
archivePrefix = {arXiv},
       eprint = {2405.20035},
 primaryClass = {astro-ph.EP},
       adsurl = {https://ui.adsabs.harvard.edu/abs/2024ApJ...968L...2L}
}

@ARTICLE{Zhou2019,
       author = {{Zhou}, G. and {Bakos}, G. {\'A}. and {Bayliss}, D. and {Bento}, J. and {Bhatti}, W. and {Brahm}, R. and {Csubry}, Z. and {Espinoza}, N. and {Hartman}, J.~D. and {Henning}, T. and {Jord{\'a}n}, A. and {Mancini}, L. and {Penev}, K. and {Rabus}, M. and {Sarkis}, P. and {Suc}, V. and {de Val-Borro}, M. and {Rodriguez}, J.~E. and {Osip}, D. and {Kedziora-Chudczer}, L. and {Bailey}, J. and {Tinney}, C.~G. and {Durkan}, S. and {L{\'a}z{\'a}r}, J. and {Papp}, I. and {S{\'a}ri}, P.},
        title = "{HATS-70b: A 13 MJ Brown Dwarf Transiting an A Star}",
      journal = {\aj},
         year = 2019,
        month = jan,
       volume = {157},
       number = {1},
          eid = {31},
        pages = {31},
          doi = {10.3847/1538-3881/aaf1bb},
archivePrefix = {arXiv},
       eprint = {1811.06925},
 primaryClass = {astro-ph.EP},
       adsurl = {https://ui.adsabs.harvard.edu/abs/2019AJ....157...31Z}
}

@ARTICLE{Meibom2007,
       author = {{Meibom}, S{\o}ren and {Mathieu}, Robert D. and {Stassun}, Keivan G.},
        title = "{The Effect of Binarity on Stellar Rotation: Beyond the Reach of Tides}",
      journal = {\apjl},
         year = 2007,
        month = aug,
       volume = {665},
       number = {2},
        pages = {L155-L158},
          doi = {10.1086/521437},
archivePrefix = {arXiv},
       eprint = {0707.1087},
 primaryClass = {astro-ph},
       adsurl = {https://ui.adsabs.harvard.edu/abs/2007ApJ...665L.155M}
}

@ARTICLE{Dai2023,
       author = {{Dai}, Fei and {Masuda}, Kento and {Beard}, Corey and {Robertson}, Paul and {Goldberg}, Max and {Batygin}, Konstantin and {Bouma}, Luke and {Lissauer}, Jack J. and {Knudstrup}, Emil and {Albrecht}, Simon and {Howard}, Andrew W. and {Knutson}, Heather A. and {Petigura}, Erik A. and {Weiss}, Lauren M. and {Isaacson}, Howard and {Kristiansen}, Martti Holst and {Osborn}, Hugh and {Wang}, Songhu and {Wang}, Xian-Yu and {Behmard}, Aida and {Greklek-McKeon}, Michael and {Vissapragada}, Shreyas and {Batalha}, Natalie M. and {Brinkman}, Casey L. and {Chontos}, Ashley and {Crossfield}, Ian and {Dressing}, Courtney and {Fetherolf}, Tara and {Fulton}, Benjamin and {Hill}, Michelle L. and {Huber}, Daniel and {Kane}, Stephen R. and {Lubin}, Jack and {MacDougall}, Mason and {Mayo}, Andrew and {Mo{\v{c}}nik}, Teo and {Akana Murphy}, Joseph M. and {Rubenzahl}, Ryan A. and {Scarsdale}, Nicholas and {Tyler}, Dakotah and {Zandt}, Judah Van and {Polanski}, Alex S. and {Schwengeler}, Hans Martin and {Terentev}, Ivan A. and {Benni}, Paul and {Bieryla}, Allyson and {Ciardi}, David and {Falk}, Ben and {Furlan}, E. and {Girardin}, Eric and {Guerra}, Pere and {Hesse}, Katharine M. and {Howell}, Steve B. and {Lillo-Box}, J. and {Matthews}, Elisabeth C. and {Twicken}, Joseph D. and {Villase{\~n}or}, Joel and {Latham}, David W. and {Jenkins}, Jon M. and {Ricker}, George R. and {Seager}, Sara and {Vanderspek}, Roland and {Winn}, Joshua N.},
        title = "{TOI-1136 is a Young, Coplanar, Aligned Planetary System in a Pristine Resonant Chain}",
      journal = {\aj},
         year = 2023,
        month = feb,
       volume = {165},
       number = {2},
          eid = {33},
        pages = {33},
          doi = {10.3847/1538-3881/aca327},
archivePrefix = {arXiv},
       eprint = {2210.09283},
 primaryClass = {astro-ph.EP},
       adsurl = {https://ui.adsabs.harvard.edu/abs/2023AJ....165...33D}
}

@ARTICLE{Zhou2018,
       author = {{Zhou}, George and {Rodriguez}, Joseph E. and {Vanderburg}, Andrew and {Quinn}, Samuel N. and {Irwin}, Jonathan and {Huang}, Chelsea X. and {Latham}, David W. and {Bieryla}, Allyson and {Esquerdo}, Gilbert A. and {Berlind}, Perry and {Calkins}, Michael L.},
        title = "{The Warm Neptunes around HD 106315 Have Low Stellar Obliquities}",
      journal = {\aj},
         year = 2018,
        month = sep,
       volume = {156},
       number = {3},
          eid = {93},
        pages = {93},
          doi = {10.3847/1538-3881/aad085},
archivePrefix = {arXiv},
       eprint = {1807.00024},
 primaryClass = {astro-ph.EP},
       adsurl = {https://ui.adsabs.harvard.edu/abs/2018AJ....156...93Z}
}

@article{Knudstrup2024,
 adsurl = {https://ui.adsabs.harvard.edu/abs/2024A&A...690A.379K},
 archiveprefix = {arXiv},
 author = {{Knudstrup}, E. and {Albrecht}, S.~H. and {Winn}, J.~N. and {Gandolfi}, D. and {Zanazzi}, J.~J. and {Persson}, C.~M. and {Fridlund}, M. and {Marcussen}, M.~L. and {Chontos}, A. and {Keniger}, M.~A.~F. and {Eisner}, N.~L. and {Bieryla}, A. and {Isaacson}, H. and {Howard}, A.~W. and {Hirsch}, L.~A. and {Murgas}, F. and {Narita}, N. and {Palle}, E. and {Kawai}, Y. and {Baker}, D.},
 doi = {10.1051/0004-6361/202450627},
 eid = {A379},
 eprint = {2408.09793},
 journal = {\aap},
 month = {October},
 pages = {A379},
 primaryclass = {astro-ph.EP},
 title = {{Obliquities of exoplanet host stars: Nineteen new and updated measurements, and trends in the sample of 205 measurements}},
 volume = {690},
 year = {2024}
}

@INPROCEEDINGS{Albrecht2007,
       author = {{Albrecht}, S. and {Reffert}, S. and {Quirrenbach}, A. and {Mitchell}, D.~S. and {Snellen}, I.},
        title = "{The Rossiter-McLaughlin Effect in the Eclipsing Binary System V1143 Cyg -- First Results}",
    booktitle = {Solar and Stellar Physics Through Eclipses},
         year = 2007,
       editor = {{Demircan}, O. and {Selam}, S.~O. and {Albayrak}, B.},
       series = {Astronomical Society of the Pacific Conference Series},
       volume = {370},
        month = may,
        pages = {218},
       adsurl = {https://ui.adsabs.harvard.edu/abs/2007ASPC..370..218A}
}

@ARTICLE{Collier2010,
       author = {{Collier Cameron}, A. and {Bruce}, V.~A. and {Miller}, G.~R.~M. and {Triaud}, A.~H.~M.~J. and {Queloz}, D.},
        title = "{Line-profile tomography of exoplanet transits - I. The Doppler shadow of HD 189733b}",
      journal = {\mnras},
         year = 2010,
        month = mar,
       volume = {403},
       number = {1},
        pages = {151-158},
          doi = {10.1111/j.1365-2966.2009.16131.x},
archivePrefix = {arXiv},
       eprint = {0911.5361},
 primaryClass = {astro-ph.SR},
       adsurl = {https://ui.adsabs.harvard.edu/abs/2010MNRAS.403..151C}
}

@ARTICLE{Radzom2024,
       author = {{Radzom}, Brandon T. and {Dong}, Jiayin and {Rice}, Malena and {Wang}, Xian-Yu and {Yee}, Samuel W. and {Fairnington}, Tyler R. and {Petrovich}, Cristobal and {Wang}, Songhu},
        title = "{Evidence for Primordial Alignment: Insights from Stellar Obliquity Measurements for Compact Sub-Saturn Systems}",
      journal = {\aj},
         year = 2024,
        month = sep,
       volume = {168},
       number = {3},
          eid = {116},
        pages = {116},
          doi = {10.3847/1538-3881/ad61d8},
archivePrefix = {arXiv},
       eprint = {2404.06504},
 primaryClass = {astro-ph.EP},
       adsurl = {https://ui.adsabs.harvard.edu/abs/2024AJ....168..116R}
}

@ARTICLE{Ferreira2024,
       author = {{Ferreira dos Santos}, Thiago and {Rice}, Malena and {Wang}, Xian-Yu and {Wang}, Songhu},
        title = "{SOLES XII. The Aligned Orbit of TOI-2533 b, a Transiting Brown Dwarf Orbiting an F8-type Star}",
      journal = {\aj},
         year = 2024,
        month = oct,
       volume = {168},
       number = {4},
          eid = {145},
        pages = {145},
          doi = {10.3847/1538-3881/ad6b7f},
archivePrefix = {arXiv},
       eprint = {2408.00725},
 primaryClass = {astro-ph.EP},
       adsurl = {https://ui.adsabs.harvard.edu/abs/2024AJ....168..145F}
}

@ARTICLE{Radzom2025,
       author = {{Radzom}, Brandon T. and {Dong}, Jiayin and {Rice}, Malena and {Wang}, Xian-Yu and {Hixenbaugh}, Kyle and {Zhou}, George and {Huang}, Chelsea X. and {Wang}, Songhu},
        title = "{Evidence for Primordial Alignment II: Insights from Stellar Obliquity Measurements for Hot Jupiters in Compact Multiplanet Systems}",
      journal = {\aj},
         year = 2025,
        month = mar,
       volume = {169},
       number = {3},
          eid = {189},
        pages = {189},
          doi = {10.3847/1538-3881/ad9dd5},
archivePrefix = {arXiv},
       eprint = {2503.03745},
 primaryClass = {astro-ph.EP},
       adsurl = {https://ui.adsabs.harvard.edu/abs/2025AJ....169..189R}
}

@ARTICLE{Rusznak2025,
       author = {{Rusznak}, Jace and {Wang}, Xian-Yu and {Rice}, Malena and {Wang}, Songhu},
        title = "{From Misaligned Sub-Saturns to Aligned Brown Dwarfs: The Highest M$_{p}$/M$_{*}$ Systems Exhibit Low Obliquities, Even around Hot Stars}",
      journal = {\apjl},
         year = 2025,
        month = apr,
       volume = {983},
       number = {2},
          eid = {L42},
        pages = {L42},
          doi = {10.3847/2041-8213/adc129},
archivePrefix = {arXiv},
       eprint = {2412.04438},
 primaryClass = {astro-ph.EP},
       adsurl = {https://ui.adsabs.harvard.edu/abs/2025ApJ...983L..42R}
}

@article{kraft1967break,
       author = {{Kraft}, Robert P.},
        title = "{Studies of Stellar Rotation. V. The Dependence of Rotation on Age among Solar-Type Stars}",
      journal = {\apj},
         year = 1967,
        month = nov,
       volume = {150},
        pages = {551},
          doi = {10.1086/149359},
       adsurl = {https://ui.adsabs.harvard.edu/abs/1967ApJ...150..551K}
}

@INCOLLECTION{Triaud2018,
       author = {{Triaud}, Amaury H.~M.~J.},
        title = "{The Rossiter-McLaughlin Effect in Exoplanet Research}",
    booktitle = {Handbook of Exoplanets},
         year = 2018,
       editor = {{Deeg}, Hans J. and {Belmonte}, Juan Antonio},
          eid = {2},
        pages = {2},
          doi = {10.1007/978-3-319-55333-7_2},
       adsurl = {https://ui.adsabs.harvard.edu/abs/2018haex.bookE...2T}
}

@ARTICLE{GaiaCollaboration2023,
       author = {{Gaia Collaboration} and {Vallenari}, A. and {Brown}, A.~G.~A. and {Prusti}, T. and {de Bruijne}, J.~H.~J. and {Arenou}, F. and {Babusiaux}, C. and {Biermann}, M. and {Creevey}, O.~L. and {Ducourant}, C. and {Evans}, D.~W. and {Eyer}, L. and {Guerra}, R. and {Hutton}, A. and {Jordi}, C. and {Klioner}, S.~A. and {Lammers}, U.~L. and {Lindegren}, L. and {Luri}, X. and {Mignard}, F. and {Panem}, C. and {Pourbaix}, D. and {Randich}, S. and {Sartoretti}, P. and {Soubiran}, C. and {Tanga}, P. and {Walton}, N.~A. and {Bailer-Jones}, C.~A.~L. and {Bastian}, U. and {Drimmel}, R. and {Jansen}, F. and {Katz}, D. and {Lattanzi}, M.~G. and {van Leeuwen}, F. and {Bakker}, J. and {Cacciari}, C. and {Casta{\~n}eda}, J. and {De Angeli}, F. and {Fabricius}, C. and {Fouesneau}, M. and {Fr{\'e}mat}, Y. and {Galluccio}, L. and {Guerrier}, A. and {Heiter}, U. and {Masana}, E. and {Messineo}, R. and {Mowlavi}, N. and {Nicolas}, C. and {Nienartowicz}, K. and {Pailler}, F. and {Panuzzo}, P. and {Riclet}, F. and {Roux}, W. and {Seabroke}, G.~M. and {Sordo}, R. and {Th{\'e}venin}, F. and {Gracia-Abril}, G. and {Portell}, J. and {Teyssier}, D. and {Altmann}, M. and {Andrae}, R. and {Audard}, M. and {Bellas-Velidis}, I. and {Benson}, K. and {Berthier}, J. and {Blomme}, R. and {Burgess}, P.~W. and {Busonero}, D. and {Busso}, G. and {C{\'a}novas}, H. and {Carry}, B. and {Cellino}, A. and {Cheek}, N. and {Clementini}, G. and {Damerdji}, Y. and {Davidson}, M. and {de Teodoro}, P. and {Nu{\~n}ez Campos}, M. and {Delchambre}, L. and {Dell'Oro}, A. and {Esquej}, P. and {Fern{\'a}ndez-Hern{\'a}ndez}, J. and {Fraile}, E. and {Garabato}, D. and {Garc{\'\i}a-Lario}, P. and {Gosset}, E. and {Haigron}, R. and {Halbwachs}, J. -L. and {Hambly}, N.~C. and {Harrison}, D.~L. and {Hern{\'a}ndez}, J. and {Hestroffer}, D. and {Hodgkin}, S.~T. and {Holl}, B. and {Jan{\ss}en}, K. and {Jevardat de Fombelle}, G. and {Jordan}, S. and {Krone-Martins}, A. and {Lanzafame}, A.~C. and {L{\"o}ffler}, W. and {Marchal}, O. and {Marrese}, P.~M. and {Moitinho}, A. and {Muinonen}, K. and {Osborne}, P. and {Pancino}, E. and {Pauwels}, T. and {Recio-Blanco}, A. and {Reyl{\'e}}, C. and {Riello}, M. and {Rimoldini}, L. and {Roegiers}, T. and {Rybizki}, J. and {Sarro}, L.~M. and {Siopis}, C. and {Smith}, M. and {Sozzetti}, A. and {Utrilla}, E. and {van Leeuwen}, M. and {Abbas}, U. and {{\'A}brah{\'a}m}, P. and {Abreu Aramburu}, A. and {Aerts}, C. and {Aguado}, J.~J. and {Ajaj}, M. and {Aldea-Montero}, F. and {Altavilla}, G. and {{\'A}lvarez}, M.~A. and {Alves}, J. and {Anders}, F. and {Anderson}, R.~I. and {Anglada Varela}, E. and {Antoja}, T. and {Baines}, D. and {Baker}, S.~G. and {Balaguer-N{\'u}{\~n}ez}, L. and {Balbinot}, E. and {Balog}, Z. and {Barache}, C. and {Barbato}, D. and {Barros}, M. and {Barstow}, M.~A. and {Bartolom{\'e}}, S. and {Bassilana}, J. -L. and {Bauchet}, N. and {Becciani}, U. and {Bellazzini}, M. and {Berihuete}, A. and {Bernet}, M. and {Bertone}, S. and {Bianchi}, L. and {Binnenfeld}, A. and {Blanco-Cuaresma}, S. and {Blazere}, A. and {Boch}, T. and {Bombrun}, A. and {Bossini}, D. and {Bouquillon}, S. and {Bragaglia}, A. and {Bramante}, L. and {Breedt}, E. and {Bressan}, A. and {Brouillet}, N. and {Brugaletta}, E. and {Bucciarelli}, B. and {Burlacu}, A. and {Butkevich}, A.~G. and {Buzzi}, R. and {Caffau}, E. and {Cancelliere}, R. and {Cantat-Gaudin}, T. and {Carballo}, R. and {Carlucci}, T. and {Carnerero}, M.~I. and {Carrasco}, J.~M. and {Casamiquela}, L. and {Castellani}, M. and {Castro-Ginard}, A. and {Chaoul}, L. and {Charlot}, P. and {Chemin}, L. and {Chiaramida}, V. and {Chiavassa}, A. and {Chornay}, N. and {Comoretto}, G. and {Contursi}, G. and {Cooper}, W.~J. and {Cornez}, T. and {Cowell}, S. and {Crifo}, F. and {Cropper}, M. and {Crosta}, M. and {Crowley}, C. and {Dafonte}, C. and {Dapergolas}, A. and {David}, M. and {David}, P. and {de Laverny}, P. and {De Luise}, F. and {De March}, R. and {De Ridder}, J. and {de Souza}, R. and {de Torres}, A. and {del Peloso}, E.~F. and {del Pozo}, E. and {Delbo}, M. and {Delgado}, A. and {Delisle}, J. -B. and {Demouchy}, C. and {Dharmawardena}, T.~E. and {Di Matteo}, P. and {Diakite}, S. and {Diener}, C. and {Distefano}, E. and {Dolding}, C. and {Edvardsson}, B. and {Enke}, H. and {Fabre}, C. and {Fabrizio}, M. and {Faigler}, S. and {Fedorets}, G. and {Fernique}, P. and {Fienga}, A. and {Figueras}, F. and {Fournier}, Y. and {Fouron}, C. and {Fragkoudi}, F. and {Gai}, M. and {Garcia-Gutierrez}, A. and {Garcia-Reinaldos}, M. and {Garc{\'\i}a-Torres}, M. and {Garofalo}, A. and {Gavel}, A. and {Gavras}, P. and {Gerlach}, E. and {Geyer}, R. and {Giacobbe}, P. and {Gilmore}, G. and {Girona}, S. and {Giuffrida}, G. and {Gomel}, R. and {Gomez}, A. and {Gonz{\'a}lez-N{\'u}{\~n}ez}, J. and {Gonz{\'a}lez-Santamar{\'\i}a}, I. and {Gonz{\'a}lez-Vidal}, J.~J. and {Granvik}, M. and {Guillout}, P. and {Guiraud}, J. and {Guti{\'e}rrez-S{\'a}nchez}, R. and {Guy}, L.~P. and {Hatzidimitriou}, D. and {Hauser}, M. and {Haywood}, M. and {Helmer}, A. and {Helmi}, A. and {Sarmiento}, M.~H. and {Hidalgo}, S.~L. and {Hilger}, T. and {H{\l}adczuk}, N. and {Hobbs}, D. and {Holland}, G. and {Huckle}, H.~E. and {Jardine}, K. and {Jasniewicz}, G. and {Jean-Antoine Piccolo}, A. and {Jim{\'e}nez-Arranz}, {\'O}. and {Jorissen}, A. and {Juaristi Campillo}, J. and {Julbe}, F. and {Karbevska}, L. and {Kervella}, P. and {Khanna}, S. and {Kontizas}, M. and {Kordopatis}, G. and {Korn}, A.~J. and {K{\'o}sp{\'a}l}, {\'A}. and {Kostrzewa-Rutkowska}, Z. and {Kruszy{\'n}ska}, K. and {Kun}, M. and {Laizeau}, P. and {Lambert}, S. and {Lanza}, A.~F. and {Lasne}, Y. and {Le Campion}, J. -F. and {Lebreton}, Y. and {Lebzelter}, T. and {Leccia}, S. and {Leclerc}, N. and {Lecoeur-Taibi}, I. and {Liao}, S. and {Licata}, E.~L. and {Lindstr{\o}m}, H.~E.~P. and {Lister}, T.~A. and {Livanou}, E. and {Lobel}, A. and {Lorca}, A. and {Loup}, C. and {Madrero Pardo}, P. and {Magdaleno Romeo}, A. and {Managau}, S. and {Mann}, R.~G. and {Manteiga}, M. and {Marchant}, J.~M. and {Marconi}, M. and {Marcos}, J. and {Marcos Santos}, M.~M.~S. and {Mar{\'\i}n Pina}, D. and {Marinoni}, S. and {Marocco}, F. and {Marshall}, D.~J. and {Martin Polo}, L. and {Mart{\'\i}n-Fleitas}, J.~M. and {Marton}, G. and {Mary}, N. and {Masip}, A. and {Massari}, D. and {Mastrobuono-Battisti}, A. and {Mazeh}, T. and {McMillan}, P.~J. and {Messina}, S. and {Michalik}, D. and {Millar}, N.~R. and {Mints}, A. and {Molina}, D. and {Molinaro}, R. and {Moln{\'a}r}, L. and {Monari}, G. and {Mongui{\'o}}, M. and {Montegriffo}, P. and {Montero}, A. and {Mor}, R. and {Mora}, A. and {Morbidelli}, R. and {Morel}, T. and {Morris}, D. and {Muraveva}, T. and {Murphy}, C.~P. and {Musella}, I. and {Nagy}, Z. and {Noval}, L. and {Oca{\~n}a}, F. and {Ogden}, A. and {Ordenovic}, C. and {Osinde}, J.~O. and {Pagani}, C. and {Pagano}, I. and {Palaversa}, L. and {Palicio}, P.~A. and {Pallas-Quintela}, L. and {Panahi}, A. and {Payne-Wardenaar}, S. and {Pe{\~n}alosa Esteller}, X. and {Penttil{\"a}}, A. and {Pichon}, B. and {Piersimoni}, A.~M. and {Pineau}, F. -X. and {Plachy}, E. and {Plum}, G. and {Poggio}, E. and {Pr{\v{s}}a}, A. and {Pulone}, L. and {Racero}, E. and {Ragaini}, S. and {Rainer}, M. and {Raiteri}, C.~M. and {Rambaux}, N. and {Ramos}, P. and {Ramos-Lerate}, M. and {Re Fiorentin}, P. and {Regibo}, S. and {Richards}, P.~J. and {Rios Diaz}, C. and {Ripepi}, V. and {Riva}, A. and {Rix}, H. -W. and {Rixon}, G. and {Robichon}, N. and {Robin}, A.~C. and {Robin}, C. and {Roelens}, M. and {Rogues}, H.~R.~O. and {Rohrbasser}, L. and {Romero-G{\'o}mez}, M. and {Rowell}, N. and {Royer}, F. and {Ruz Mieres}, D. and {Rybicki}, K.~A. and {Sadowski}, G. and {S{\'a}ez N{\'u}{\~n}ez}, A. and {Sagrist{\`a} Sell{\'e}s}, A. and {Sahlmann}, J. and {Salguero}, E. and {Samaras}, N. and {Sanchez Gimenez}, V. and {Sanna}, N. and {Santove{\~n}a}, R. and {Sarasso}, M. and {Schultheis}, M. and {Sciacca}, E. and {Segol}, M. and {Segovia}, J.~C. and {S{\'e}gransan}, D. and {Semeux}, D. and {Shahaf}, S. and {Siddiqui}, H.~I. and {Siebert}, A. and {Siltala}, L. and {Silvelo}, A. and {Slezak}, E. and {Slezak}, I. and {Smart}, R.~L. and {Snaith}, O.~N. and {Solano}, E. and {Solitro}, F. and {Souami}, D. and {Souchay}, J. and {Spagna}, A. and {Spina}, L. and {Spoto}, F. and {Steele}, I.~A. and {Steidelm{\"u}ller}, H. and {Stephenson}, C.~A. and {S{\"u}veges}, M. and {Surdej}, J. and {Szabados}, L. and {Szegedi-Elek}, E. and {Taris}, F. and {Taylor}, M.~B. and {Teixeira}, R. and {Tolomei}, L. and {Tonello}, N. and {Torra}, F. and {Torra}, J. and {Torralba Elipe}, G. and {Trabucchi}, M. and {Tsounis}, A.~T. and {Turon}, C. and {Ulla}, A. and {Unger}, N. and {Vaillant}, M.~V. and {van Dillen}, E. and {van Reeven}, W. and {Vanel}, O. and {Vecchiato}, A. and {Viala}, Y. and {Vicente}, D. and {Voutsinas}, S. and {Weiler}, M. and {Wevers}, T. and {Wyrzykowski}, {\L}. and {Yoldas}, A. and {Yvard}, P. and {Zhao}, H. and {Zorec}, J. and {Zucker}, S. and {Zwitter}, T.},
        title = "{Gaia Data Release 3. Summary of the content and survey properties}",
      journal = {\aap},
         year = 2023,
        month = jun,
       volume = {674},
          eid = {A1},
        pages = {A1},
          doi = {10.1051/0004-6361/202243940},
archivePrefix = {arXiv},
       eprint = {2208.00211},
 primaryClass = {astro-ph.GA},
       adsurl = {https://ui.adsabs.harvard.edu/abs/2023A&A...674A...1G}
}

@ARTICLE{WinnFabrycky2015,
       author = {{Winn}, Joshua N. and {Fabrycky}, Daniel C.},
        title = "{The Occurrence and Architecture of Exoplanetary Systems}",
      journal = {\araa},
         year = 2015,
        month = aug,
       volume = {53},
        pages = {409-447},
          doi = {10.1146/annurev-astro-082214-122246},
archivePrefix = {arXiv},
       eprint = {1410.4199},
 primaryClass = {astro-ph.EP},
       adsurl = {https://ui.adsabs.harvard.edu/abs/2015ARA&A..53..409W}
}

@ARTICLE{Guenther2012,
       author = {{Guenther}, E.~W. and {D{\'\i}az}, R.~F. and {Gazzano}, J. -C. and {Mazeh}, T. and {Rouan}, D. and {Gibson}, N. and {Csizmadia}, Sz. and {Aigrain}, S. and {Alonso}, R. and {Almenara}, J.~M. and {Auvergne}, M. and {Baglin}, A. and {Barge}, P. and {Bonomo}, A.~S. and {Bord{\'e}}, P. and {Bouchy}, F. and {Bruntt}, H. and {Cabrera}, J. and {Carone}, L. and {Carpano}, S. and {Cavarroc}, C. and {Deeg}, H.~J. and {Deleuil}, M. and {Dreizler}, S. and {Dvorak}, R. and {Erikson}, A. and {Ferraz-Mello}, S. and {Fridlund}, M. and {Gandolfi}, D. and {Gillon}, M. and {Guillot}, T. and {Hatzes}, A. and {Havel}, M. and {H{\'e}brard}, G. and {Jehin}, E. and {Jorda}, L. and {Lammer}, H. and {L{\'e}ger}, A. and {Moutou}, C. and {Nortmann}, L. and {Ollivier}, M. and {Ofir}, A. and {Pasternacki}, Th. and {P{\"a}tzold}, M. and {Parviainen}, H. and {Queloz}, D. and {Rauer}, H. and {Samuel}, B. and {Santerne}, A. and {Schneider}, J. and {Tal-Or}, L. and {Tingley}, B. and {Weingrill}, J. and {Wuchterl}, G.},
        title = "{Transiting exoplanets from the CoRoT space mission. XXI. CoRoT-19b: a low density planet orbiting an old inactive F9V-star}",
      journal = {\aap},
         year = 2012,
        month = jan,
       volume = {537},
          eid = {A136},
        pages = {A136},
          doi = {10.1051/0004-6361/201117706},
archivePrefix = {arXiv},
       eprint = {1112.1035},
 primaryClass = {astro-ph.EP},
       adsurl = {https://ui.adsabs.harvard.edu/abs/2012A&A...537A.136G}
}

@ARTICLE{Li2016,
       author = {{Li}, Gongjie and {Winn}, Joshua N.},
        title = "{Are Tidal Effects Responsible for Exoplanetary Spin\&amp;ndashOrbit Alignment?}",
      journal = {\apj},
         year = 2016,
        month = feb,
       volume = {818},
       number = {1},
          eid = {5},
        pages = {5},
          doi = {10.3847/0004-637X/818/1/5},
archivePrefix = {arXiv},
       eprint = {1511.05570},
 primaryClass = {astro-ph.EP},
       adsurl = {https://ui.adsabs.harvard.edu/abs/2016ApJ...818....5L}
}

@ARTICLE{zanazzi2024damping,
       author = {{Zanazzi}, J.~J. and {Dewberry}, Janosz and {Chiang}, Eugene},
        title = "{Damping Obliquities of Hot Jupiter Hosts by Resonance Locking}",
      journal = {\apjl},
         year = 2024,
        month = jun,
       volume = {967},
       number = {2},
          eid = {L29},
        pages = {L29},
          doi = {10.3847/2041-8213/ad4644},
archivePrefix = {arXiv},
       eprint = {2403.05616},
 primaryClass = {astro-ph.EP},
       adsurl = {https://ui.adsabs.harvard.edu/abs/2024ApJ...967L..29Z}
}

@article{Naoz2012,
  title={On the formation of hot Jupiters in stellar binaries},
  author={Naoz, Smadar and Farr, Will M and Rasio, Frederic A},
  journal={The Astrophysical Journal Letters},
  volume={754},
  number={2},
  pages={L36},
  year={2012},
  publisher={IOP Publishing}
}

@ARTICLE{Fabrycky2007,
       author = {{Fabrycky}, Daniel and {Tremaine}, Scott},
        title = "{Shrinking Binary and Planetary Orbits by Kozai Cycles with Tidal Friction}",
      journal = {\apj},
         year = 2007,
        month = nov,
       volume = {669},
       number = {2},
        pages = {1298-1315},
          doi = {10.1086/521702},
archivePrefix = {arXiv},
       eprint = {0705.4285},
 primaryClass = {astro-ph},
       adsurl = {https://ui.adsabs.harvard.edu/abs/2007ApJ...669.1298F}
}

@ARTICLE{Wu2003,
       author = {{Wu}, Y. and {Murray}, N.},
        title = "{Planet Migration and Binary Companions: The Case of HD 80606b}",
      journal = {\apj},
         year = 2003,
        month = may,
       volume = {589},
       number = {1},
        pages = {605-614},
          doi = {10.1086/374598},
archivePrefix = {arXiv},
       eprint = {astro-ph/0303010},
 primaryClass = {astro-ph},
       adsurl = {https://ui.adsabs.harvard.edu/abs/2003ApJ...589..605W}
}

@ARTICLE{Albrecht2012,
       author = {{Albrecht}, Simon and {Winn}, Joshua N. and {Johnson}, John A. and
         {Howard}, Andrew W. and {Marcy}, Geoffrey W. and {Butler}, R. Paul and
         {Arriagada}, Pamela and {Crane}, Jeffrey D. and {Shectman}, Stephen A. and
         {Thompson}, Ian B. and {Hirano}, Teruyuki and {Bakos}, Gaspar and
         {Hartman}, Joel D.},
        title = "{Obliquities of Hot Jupiter Host Stars: Evidence for Tidal Interactions and Primordial Misalignments}",
      journal = {\apj},
         year = 2012,
        month = sep,
       volume = {757},
       number = {1},
          eid = {18},
        pages = {18},
          doi = {10.1088/0004-637X/757/1/18},
archivePrefix = {arXiv},
       eprint = {1206.6105},
 primaryClass = {astro-ph.SR},
       adsurl = {https://ui.adsabs.harvard.edu/abs/2012ApJ...757...18A}
}

@ARTICLE{Gaia,
   author = {{Gaia Collaboration} and {Brown}, A.~G.~A. and {Vallenari}, A. and 
	{Prusti}, T. and {de Bruijne}, J.~H.~J. and {Babusiaux}, C. and 
	{Bailer-Jones}, C.~A.~L.},
    title = "{Gaia Data Release 2. Summary of the contents and survey properties}",
  journal = {ArXiv e-prints},
archivePrefix = "arXiv",
   eprint = {1804.09365},
     year = 2018,
    month = apr,
   adsurl = {http://adsabs.harvard.edu/abs/2018arXiv180409365G}
}

@ARTICLE{Johnson2010,
       author = {{Johnson}, John Asher and {Aller}, Kimberly M. and {Howard}, Andrew W. and
         {Crepp}, Justin R.},
        title = "{Giant Planet Occurrence in the Stellar Mass-Metallicity Plane}",
      journal = {\pasp},
         year = "2010",
        month = "Aug",
       volume = {122},
       number = {894},
        pages = {905},
          doi = {10.1086/655775},
archivePrefix = {arXiv},
       eprint = {1005.3084},
 primaryClass = {astro-ph.EP},
       adsurl = {https://ui.adsabs.harvard.edu/abs/2010PASP..122..905J}
}

@dataset{Cutri2014,
       author = {{Cutri}, R.~M. and {Wright}, E.~L. and {Conrow}, T. and {Fowler}, J.~W. and {Eisenhardt}, P.~R.~M. and {Grillmair}, C. and {Kirkpatrick}, J.~D. and {Masci}, F. and {McCallon}, H.~L. and {Wheelock}, S.~L. and {Fajardo-Acosta}, S. and {Yan}, L. and {Benford}, D. and {Harbut}, M. and {Jarrett}, T. and {Lake}, S. and {Leisawitz}, D. and {Ressler}, M.~E. and {Stanford}, S.~A. and {Tsai}, C.-W. and {Liu}, F. and {Helou}, G. and {Mainzer}, A. and {Gettngs}, D. and {Gonzalez}, A. and {Hoffman}, D. and {Marsh}, K.~A. and {Padgett}, D. and {Skrutskie}, M.~F. and {Beck}, R. and {Papin}, M. and {Wittman}, M.},
        title = "{VizieR Online Data Catalog: AllWISE Data Release (Cutri+ 2013)}",
 howpublished = {VizieR On-line Data Catalog: II/328.  Originally published in: IPAC/Caltech (2013)},
         year = 2021,
        month = feb,
          eid = {II/328},
       adsurl = {https://ui.adsabs.harvard.edu/abs/2014yCat.2328....0C}
}

@ARTICLE{Santos2013sweetcat,
       author = {{Santos}, N.~C. and {Sousa}, S.~G. and {Mortier}, A. and {Neves}, V. and {Adibekyan}, V. and {Tsantaki}, M. and {Delgado Mena}, E. and {Bonfils}, X. and {Israelian}, G. and {Mayor}, M. and {Udry}, S.},
        title = "{SWEET-Cat: A catalogue of parameters for Stars With ExoplanETs. I. New atmospheric parameters and masses for 48 stars with planets}",
      journal = {\aap},
         year = 2013,
        month = aug,
       volume = {556},
          eid = {A150},
        pages = {A150},
          doi = {10.1051/0004-6361/201321286},
archivePrefix = {arXiv},
       eprint = {1307.0354},
 primaryClass = {astro-ph.SR},
       adsurl = {https://ui.adsabs.harvard.edu/abs/2013A&A...556A.150S}
}

@article{Sousa2021,
 adsurl = {https://ui.adsabs.harvard.edu/abs/2021A&A...656A..53S},
 archiveprefix = {arXiv},
 author = {{Sousa}, S.~G. and {Adibekyan}, V. and {Delgado-Mena}, E. and {Santos}, N.~C. and {Rojas-Ayala}, B. and {Soares}, B.~M.~T.~B. and {Legoinha}, H. and {Ulmer-Moll}, S. and {Camacho}, J.~D. and {Barros}, S.~C.~C. and {Demangeon}, O.~D.~S. and {Hoyer}, S. and {Israelian}, G. and {Mortier}, A. and {Tsantaki}, M. and {Monteiro}, M.~A.},
 doi = {10.1051/0004-6361/202141584},
 eid = {A53},
 eprint = {2109.04781},
 journal = {\aap},
 month = {December},
 pages = {A53},
 primaryclass = {astro-ph.EP},
 title = {{SWEET-Cat 2.0: The Cat just got SWEETer. Higher quality spectra and precise parallaxes from Gaia eDR3}},
 volume = {656},
 year = {2021}
}

@ARTICLE{Dugan2025,
       author = {{Dugan}, Emma and {Wang}, Xian-Yu and {Heron}, Agustin and {Gautham Bhaskar}, Hareesh and {Rice}, Malena and {Petrovich}, Cristobal and {Wang}, Songhu},
        title = "{Early Evidence for Polar Orbits of Sub-Saturns Around Hot Stars}",
      journal = {arXiv e-prints},
         year = 2025,
        month = oct,
          eid = {arXiv:2510.20740},
        pages = {arXiv:2510.20740},
          doi = {10.48550/arXiv.2510.20740},
archivePrefix = {arXiv},
       eprint = {2510.20740},
 primaryClass = {astro-ph.EP},
       adsurl = {https://ui.adsabs.harvard.edu/abs/2025arXiv251020740D}
}

@ARTICLE{Fu2021,
       author = {{Fu}, Guangwei and {Deming}, Drake and {Lothringer}, Joshua and {Nikolov}, Nikolay and {Sing}, David K. and {Kempton}, Eliza M.-R. and {Ih}, Jegug and {Evans}, Thomas M. and {Stevenson}, Kevin and {Wakeford}, H.~R. and {Rodriguez}, Joseph E. and {Eastman}, Jason D. and {Stassun}, Keivan and {Henry}, Gregory W. and {L{\'o}pez-Morales}, Mercedes and {Lendl}, Monika and {Conti}, Dennis M. and {Stockdale}, Chris and {Collins}, Karen and {Kielkopf}, John and {Barstow}, Joanna K. and {Sanz-Forcada}, Jorge and {Ehrenreich}, David and {Bourrier}, Vincent and {dos Santos}, Leonardo A.},
        title = "{The Hubble PanCET Program: Transit and Eclipse Spectroscopy of the Strongly Irradiated Giant Exoplanet WASP-76b}",
      journal = {\aj},
         year = 2021,
        month = sep,
       volume = {162},
       number = {3},
          eid = {108},
        pages = {108},
          doi = {10.3847/1538-3881/ac1200},
archivePrefix = {arXiv},
       eprint = {2005.02568},
 primaryClass = {astro-ph.EP},
       adsurl = {https://ui.adsabs.harvard.edu/abs/2021AJ....162..108F}
}

@ARTICLE{Stassun2017,
       author = {{Stassun}, Keivan G. and {Collins}, Karen A. and {Gaudi}, B. Scott},
        title = "{Accurate Empirical Radii and Masses of Planets and Their Host Stars with Gaia Parallaxes}",
      journal = {\aj},
         year = 2017,
        month = mar,
       volume = {153},
       number = {3},
          eid = {136},
        pages = {136},
          doi = {10.3847/1538-3881/aa5df3},
archivePrefix = {arXiv},
       eprint = {1609.04389},
 primaryClass = {astro-ph.EP},
       adsurl = {https://ui.adsabs.harvard.edu/abs/2017AJ....153..136S}
}

@ARTICLE{NeveuVanMalle2014,
       author = {{Neveu-VanMalle}, M. and {Queloz}, D. and {Anderson}, D.~R. and {Charbonnel}, C. and {Collier Cameron}, A. and {Delrez}, L. and {Gillon}, M. and {Hellier}, C. and {Jehin}, E. and {Lendl}, M. and {Maxted}, P.~F.~L. and {Pepe}, F. and {Pollacco}, D. and {S{\'e}gransan}, D. and {Smalley}, B. and {Smith}, A.~M.~S. and {Southworth}, J. and {Triaud}, A.~H.~M.~J. and {Udry}, S. and {West}, R.~G.},
        title = "{WASP-94 A and B planets: hot-Jupiter cousins in a twin-star system}",
      journal = {\aap},
         year = 2014,
        month = dec,
       volume = {572},
          eid = {A49},
        pages = {A49},
          doi = {10.1051/0004-6361/201424744},
archivePrefix = {arXiv},
       eprint = {1409.7566},
 primaryClass = {astro-ph.EP},
       adsurl = {https://ui.adsabs.harvard.edu/abs/2014A&A...572A..49N}
}

@ARTICLE{Leonardi2024,
       author = {{Leonardi}, P. and {Nascimbeni}, V. and {Granata}, V. and {Malavolta}, L. and {Borsato}, L. and {Biazzo}, K. and {Lanza}, A.~F. and {Desidera}, S. and {Piotto}, G. and {Nardiello}, D. and {Damasso}, M. and {Cunial}, A. and {Bedin}, L.~R.},
        title = "{TASTE. V. A new ground-based investigation of orbital decay in the ultra-hot Jupiter WASP-12b}",
      journal = {\aap},
         year = 2024,
        month = jun,
       volume = {686},
          eid = {A84},
        pages = {A84},
          doi = {10.1051/0004-6361/202348363},
archivePrefix = {arXiv},
       eprint = {2402.12120},
 primaryClass = {astro-ph.EP},
       adsurl = {https://ui.adsabs.harvard.edu/abs/2024A&A...686A..84L}
}

@ARTICLE{Johns2019,
       author = {{Johns}, Daniel and {Reed}, Phillip A. and {Rodriguez}, Joseph E. and {Pepper}, Joshua and {Stassun}, Keivan G. and {Penev}, Kaloyan and {Gaudi}, B. Scott and {Labadie-Bartz}, Jonathan and {Fulton}, Benjamin J. and {Quinn}, Samuel N. and {Eastman}, Jason D. and {Ciardi}, David R. and {Hirsch}, Lea and {Stevens}, Daniel J. and {Stevens}, Catherine P. and {Oberst}, Thomas E. and {Cohen}, David H. and {Jensen}, Eric L.~N. and {Benni}, Paul and {Villanueva}, Jr., Steven and {Murawski}, Gabriel and {Bieryla}, Allyson and {Latham}, David W. and {Vanaverbeke}, Siegfried and {Dubois}, Franky and {Rau}, Steve and {Logie}, Ludwig and {Rauenzahn}, Ryan F. and {Wittenmyer}, Robert A. and {Zambelli}, Roberto and {Bayliss}, Daniel and {Beatty}, Thomas G. and {Collins}, Karen A. and {Col{\'o}n}, Knicole D. and {Curtis}, Ivan A. and {Evans}, Phil and {Gregorio}, Joao and {James}, David and {Depoy}, D.~L. and {Johnson}, Marshall C. and {Joner}, Michael D. and {Kasper}, David H. and {Khakpash}, Somayeh and {Kielkopf}, John F. and {Kuhn}, Rudolf B. and {Lund}, Michael B. and {Manner}, Mark and {Marshall}, Jennifer L. and {McLeod}, Kim K. and {Penny}, Matthew T. and {Relles}, Howard and {Siverd}, Robert J. and {Stephens}, Denise C. and {Stockdale}, Chris and {Tan}, Thiam-Guan and {Trueblood}, Mark and {Trueblood}, Pat and {Yao}, Xinyu},
        title = "{KELT-23Ab: A Hot Jupiter Transiting a Near-solar Twin Close to the TESS and JWST Continuous Viewing Zones}",
      journal = {\aj},
         year = 2019,
        month = aug,
       volume = {158},
       number = {2},
          eid = {78},
        pages = {78},
          doi = {10.3847/1538-3881/ab24c7},
archivePrefix = {arXiv},
       eprint = {1903.00031},
 primaryClass = {astro-ph.EP},
       adsurl = {https://ui.adsabs.harvard.edu/abs/2019AJ....158...78J}
}

@ARTICLE{Eastman2016,
       author = {{Eastman}, Jason D. and {Beatty}, Thomas G. and {Siverd}, Robert J. and {Antognini}, Joseph M.~O. and {Penny}, Matthew T. and {Gonzales}, Erica J. and {Crepp}, Justin R. and {Howard}, Andrew W. and {Avril}, Ryan L. and {Bieryla}, Allyson and {Collins}, Karen and {Fulton}, Benjamin J. and {Ge}, Jian and {Gregorio}, Joao and {Ma}, Bo and {Mellon}, Samuel N. and {Oberst}, Thomas E. and {Wang}, Ji and {Gaudi}, B. Scott and {Pepper}, Joshua and {Stassun}, Keivan G. and {Buchhave}, Lars A. and {Jensen}, Eric L.~N. and {Latham}, David W. and {Berlind}, Perry and {Calkins}, Michael L. and {Cargile}, Phillip A. and {Col{\'o}n}, Knicole D. and {Dhital}, Saurav and {Esquerdo}, Gilbert A. and {Johnson}, John Asher and {Kielkopf}, John F. and {Manner}, Mark and {Mao}, Qingqing and {McLeod}, Kim K. and {Penev}, Kaloyan and {Stefanik}, Robert P. and {Street}, Rachel and {Zambelli}, Roberto and {DePoy}, D.~L. and {Gould}, Andrew and {Marshall}, Jennifer L. and {Pogge}, Richard W. and {Trueblood}, Mark and {Trueblood}, Patricia},
        title = "{KELT-4Ab: An Inflated Hot Jupiter Transiting the Bright (V {\ensuremath{\sim}} 10) Component of a Hierarchical Triple}",
      journal = {\aj},
         year = 2016,
        month = feb,
       volume = {151},
       number = {2},
          eid = {45},
        pages = {45},
          doi = {10.3847/0004-6256/151/2/45},
archivePrefix = {arXiv},
       eprint = {1510.00015},
 primaryClass = {astro-ph.EP},
       adsurl = {https://ui.adsabs.harvard.edu/abs/2016AJ....151...45E}
}

@ARTICLE{Smith2019,
       author = {{Smith}, A.~M.~S. and {Csizmadia}, Sz. and {Gandolfi}, D. and {Albrecht}, S. and {Alonso}, R. and {Barrag{\'a}n}, O. and {Cabrera}, J. and {Cochran}, W.~D. and {Dai}, F. and {Deeg}, H. and {Eigm{\"u}ller}, Ph. and {Endl}, M. and {Erikson}, A. and {Fridlund}, M. and {Fukui}, A. and {Grziwa}, S. and {Guenther}, E.~W. and {Hatzes}, A.~P. and {Hidalgo}, D. and {Hirano}, T. and {Korth}, J. and {Kuzuhara}, M. and {Livingston}, J. and {Narita}, N. and {Nespral}, D. and {Niraula}, P. and {Nowak}, G. and {Palle}, E. and {P{\"a}tzold}, M. and {Persson}, C.~M. and {Prieto-Arranz}, J. and {Rauer}, H. and {Redfield}, S. and {Ribas}, I. and {Van Eylen}, V.},
        title = "{K2-295 b and K2-237 b: Two Transiting Hot Jupiters}",
      journal = {\actaa},
         year = 2019,
        month = jun,
       volume = {69},
       number = {2},
        pages = {135-158},
          doi = {10.32023/0001-5237/69.2.3},
archivePrefix = {arXiv},
       eprint = {1807.05865},
 primaryClass = {astro-ph.EP},
       adsurl = {https://ui.adsabs.harvard.edu/abs/2019AcA....69..135S}
}

@ARTICLE{Zhao2014,
       author = {{Zhao}, Ming and {O'Rourke}, Joseph G. and {Wright}, Jason T. and {Knutson}, Heather A. and {Burrows}, Adam and {Fortney}, Johnathan and {Ngo}, Henry and {Fulton}, Benjamin J. and {Baranec}, Christoph and {Riddle}, Reed and {Law}, Nicholas M. and {Muirhead}, Philip S. and {Hinkley}, Sasha and {Showman}, Adam P. and {Curtis}, Jason and {Burruss}, Rick},
        title = "{Characterization of the Atmosphere of the Hot Jupiter HAT-P-32Ab and the M-dwarf Companion HAT-P-32B}",
      journal = {\apj},
         year = 2014,
        month = dec,
       volume = {796},
       number = {2},
          eid = {115},
        pages = {115},
          doi = {10.1088/0004-637X/796/2/115},
archivePrefix = {arXiv},
       eprint = {1410.0968},
 primaryClass = {astro-ph.EP},
       adsurl = {https://ui.adsabs.harvard.edu/abs/2014ApJ...796..115Z}
}

@ARTICLE{Mortier2013,
       author = {{Mortier}, A. and {Santos}, N.~C. and {Sousa}, S.~G. and {Fernandes}, J.~M. and {Adibekyan}, V. Zh. and {Delgado Mena}, E. and {Montalto}, M. and {Israelian}, G.},
        title = "{New and updated stellar parameters for 90 transit hosts. The effect of the surface gravity}",
      journal = {\aap},
         year = 2013,
        month = oct,
       volume = {558},
          eid = {A106},
        pages = {A106},
          doi = {10.1051/0004-6361/201322240},
archivePrefix = {arXiv},
       eprint = {1309.1998},
 primaryClass = {astro-ph.EP},
       adsurl = {https://ui.adsabs.harvard.edu/abs/2013A&A...558A.106M}
}

@ARTICLE{Benomar2014,
       author = {{Benomar}, Othman and {Masuda}, Kento and {Shibahashi}, Hiromoto and {Suto}, Yasushi},
        title = "{Determination of three-dimensional spin-orbit angle with joint analysis of asteroseismology, transit lightcurve, and the Rossiter-McLaughlin effect: Cases of HAT-P-7 and Kepler-25}",
      journal = {\pasj},
         year = 2014,
        month = oct,
       volume = {66},
       number = {5},
          eid = {94},
        pages = {94},
          doi = {10.1093/pasj/psu069},
archivePrefix = {arXiv},
       eprint = {1407.7332},
 primaryClass = {astro-ph.SR},
       adsurl = {https://ui.adsabs.harvard.edu/abs/2014PASJ...66...94B}
}

@ARTICLE{Furlan2020,
       author = {{Furlan}, E. and {Howell}, S.~B.},
        title = "{Unresolved Binary Exoplanet Host Stars Fit as Single Stars: Effects on the Stellar Parameters}",
      journal = {\apj},
         year = 2020,
        month = jul,
       volume = {898},
       number = {1},
          eid = {47},
        pages = {47},
          doi = {10.3847/1538-4357/ab9c9c},
archivePrefix = {arXiv},
       eprint = {2006.06528},
 primaryClass = {astro-ph.SR},
       adsurl = {https://ui.adsabs.harvard.edu/abs/2020ApJ...898...47F}
}

@ARTICLE{ElBadry2018,
       author = {{El-Badry}, Kareem and {Rix}, Hans-Walter and {Ting}, Yuan-Sen and {Weisz}, Daniel R. and {Bergemann}, Maria and {Cargile}, Phillip and {Conroy}, Charlie and {Eilers}, Anna-Christina},
        title = "{Signatures of unresolved binaries in stellar spectra: implications for spectral fitting}",
      journal = {\mnras},
         year = 2018,
        month = feb,
       volume = {473},
       number = {4},
        pages = {5043-5049},
          doi = {10.1093/mnras/stx2758},
archivePrefix = {arXiv},
       eprint = {1709.03983},
 primaryClass = {astro-ph.SR},
       adsurl = {https://ui.adsabs.harvard.edu/abs/2018MNRAS.473.5043E}
}

@INPROCEEDINGS{Bouchy2009CoRoT1,
       author = {{Bouchy}, Fran{\c{c}}ois and {Moutou}, Claire and {Queloz}, Didier and {CoRoT Exoplanet Science Team}},
        title = "{Radial velocity follow-up for confirmation and characterization of transiting exoplanets}",
    booktitle = {Transiting Planets},
         year = 2009,
       editor = {{Pont}, Fr{\'e}d{\'e}ric and {Sasselov}, Dimitar and {Holman}, Matthew J.},
       series = {IAU Symposium},
       volume = {253},
        month = feb,
        pages = {129-139},
          doi = {10.1017/S174392130802632X},
archivePrefix = {arXiv},
       eprint = {0902.3520},
 primaryClass = {astro-ph.EP},
       adsurl = {https://ui.adsabs.harvard.edu/abs/2009IAUS..253..129B}
}

@misc{10.26133/NEA2,
  doi = {10.26133/NEA2},
  url = {https://catcopy.ipac.caltech.edu/dois/doi.php?id=10.26133/NEA2},
  author = {{NASA Exoplanet Archive}},
  title = {Composite Planet Data Table},
  publisher = {IPAC},
  year = {2019}
}

@ARTICLE{Lindegren2018,
       author = {{Lindegren}, L. and {Hern{\'a}ndez}, J. and {Bombrun}, A. and {Klioner}, S. and {Bastian}, U. and {Ramos-Lerate}, M. and {de Torres}, A. and {Steidelm{\"u}ller}, H. and {Stephenson}, C. and {Hobbs}, D. and {Lammers}, U. and {Biermann}, M. and {Geyer}, R. and {Hilger}, T. and {Michalik}, D. and {Stampa}, U. and {McMillan}, P.~J. and {Casta{\~n}eda}, J. and {Clotet}, M. and {Comoretto}, G. and {Davidson}, M. and {Fabricius}, C. and {Gracia}, G. and {Hambly}, N.~C. and {Hutton}, A. and {Mora}, A. and {Portell}, J. and {van Leeuwen}, F. and {Abbas}, U. and {Abreu}, A. and {Altmann}, M. and {Andrei}, A. and {Anglada}, E. and {Balaguer-N{\'u}{\~n}ez}, L. and {Barache}, C. and {Becciani}, U. and {Bertone}, S. and {Bianchi}, L. and {Bouquillon}, S. and {Bourda}, G. and {Br{\"u}semeister}, T. and {Bucciarelli}, B. and {Busonero}, D. and {Buzzi}, R. and {Cancelliere}, R. and {Carlucci}, T. and {Charlot}, P. and {Cheek}, N. and {Crosta}, M. and {Crowley}, C. and {de Bruijne}, J. and {de Felice}, F. and {Drimmel}, R. and {Esquej}, P. and {Fienga}, A. and {Fraile}, E. and {Gai}, M. and {Garralda}, N. and {Gonz{\'a}lez-Vidal}, J.~J. and {Guerra}, R. and {Hauser}, M. and {Hofmann}, W. and {Holl}, B. and {Jordan}, S. and {Lattanzi}, M.~G. and {Lenhardt}, H. and {Liao}, S. and {Licata}, E. and {Lister}, T. and {L{\"o}ffler}, W. and {Marchant}, J. and {Martin-Fleitas}, J. -M. and {Messineo}, R. and {Mignard}, F. and {Morbidelli}, R. and {Poggio}, E. and {Riva}, A. and {Rowell}, N. and {Salguero}, E. and {Sarasso}, M. and {Sciacca}, E. and {Siddiqui}, H. and {Smart}, R.~L. and {Spagna}, A. and {Steele}, I. and {Taris}, F. and {Torra}, J. and {van Elteren}, A. and {van Reeven}, W. and {Vecchiato}, A.},
        title = "{Gaia Data Release 2. The astrometric solution}",
      journal = {\aap},
         year = 2018,
        month = aug,
       volume = {616},
          eid = {A2},
        pages = {A2},
          doi = {10.1051/0004-6361/201832727},
archivePrefix = {arXiv},
       eprint = {1804.09366},
 primaryClass = {astro-ph.IM},
       adsurl = {https://ui.adsabs.harvard.edu/abs/2018A&A...616A...2L}
}
\bibliographystyle{aasjournal}

\end{document}